\documentstyle[aps,preprint,tighten,eqsecnum]{revtex}

\input epsf

 \newcommand \be {\begin{equation}}
\newcommand \ee {\end{equation}}
\newcommand \ba {\begin{array}{c}}
\newcommand \ea { \end{array}}
 \newcommand \bea {\begin{eqnarray}}
\newcommand \eea {\end{eqnarray}}
\def\oppropto{\mathop{\propto}}
\def\operarrow{\mathop{\longrightarrow}}
\def\opsimeq{\mathop{\simeq}}

\begin{document}

\draft

\preprint{Saclay T96/072}

\title{\bf Exponents appearing in heterogeneous reaction-diffusion models in one dimension }

\author{ C\'ecile Monthus }

\address{ Service de Physique Th\'eorique,
CEA  Saclay,  
91191 Gif-sur-Yvette, FRANCE  \\
e-mail : monthus@spht.saclay.cea.fr  }

\date{July 1996 }

\maketitle

\begin{abstract}

{
We study the following 1D two-species reaction diffusion model :
there is a small concentration of B-particles with diffusion constant $D_B$
in an homogenous background of W-particles with diffusion constant $D_W$;
 two W-particles of the majority species either coagulate ($W+W \longrightarrow W$)
 or annihilate ($W+W \longrightarrow \emptyset$) with the respective probabilities $ p_c=(q-2)/(q-1) $ and $p_a=1/(q-1)$;
 a B-particle and a W-particle annihilate ($W+B \longrightarrow \emptyset$)
 with probability $1$. 
The exponent $\theta\left(q,\lambda=D_B/D_W\right)$
describing the asymptotic time decay of the minority B-species concentration
can be viewed as a generalization of the exponent of persistent spins
in the zero-temperature Glauber dynamics of the 1D $q$-state Potts model 
starting from a random initial condition :
the W-particles represent domain walls, and the exponent $\theta(q,\lambda)$
characterizes the time decay of the probability
that a diffusive ``spectator" does not meet a domain wall up to time $t$.
We extend the methods introduced
by Derrida, Hakim and Pasquier ({\em Phys. Rev. Lett.} {\bf 75} 751 (1995);
Saclay preprint T96/013, to appear in {\em J. Stat. Phys.} (1996)) 
for the problem of persistent spins,  
to compute the exponent $\theta(q,\lambda)$ 
in perturbation at first order in $(q-1)$ for arbitrary $\lambda$ 
and at first order in $\lambda$ for arbitrary $q$.
}

\end{abstract}

\pacs{ 05.40.+j, 02.50.-r, 82.20.-w}

\vfill

\newpage

\section{Introduction}

The one-dimensional Ising or Potts model evolving according
to zero-temperature Glauber dynamics \cite{Glau63} from a random initial condition is one of the simplest
systems for which domain coarsening \cite{Bray94} can be studied
in great details. The possibility of writing closed kinetic 
equations for the expectation value of each spin and for
the equal-time two-point correlation functions \cite{Glau63,Bray90,AF90}
can be used to obtain various exact results, such as the growth in time 
of the characteristic length of the coarsening like $t^{1/2}$,
as expected in general when the order parameter is not conserved. 
More recently, it was shown that more refined quantities
such as the fraction of spins which have never flipped up to time $t$
\cite{DHP1,DHP2},
or the distribution of domain sizes \cite{DZ96} could also be studied
by mapping the problem to an exactly soluble one-species coagulation
model ($A+A \longrightarrow A$).

The zero-temperature Glauber dynamics of the $q$-state Potts model
starting from a random initial condition is related to various reaction-diffusion problems.
The simplest relation deals with the dynamics of domain walls $W$
(\cite{Der95} and references therein), that diffuse and react whenever
 they meet according to
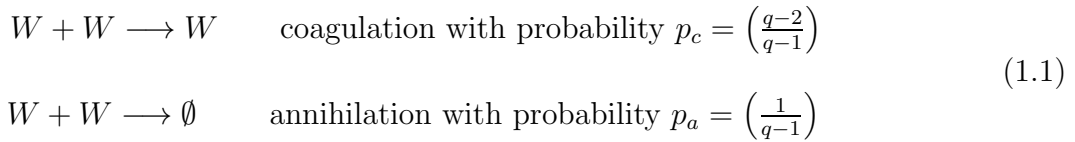
\be \ba
W+W \longrightarrow W \qquad \hbox{coagulation with probability}  \ p_c=\left({{q-2} \over {q-1}} \right) 
\\ \\
W+W \longrightarrow \emptyset \qquad \ \hbox{annihilation with probability} \ p_a=\left({{1} \over {q-1}} \right)
\label{Wrules}
\ea \ee
As such this reaction-diffusion problem has a meaning for any real value 
$0 \leq p_c \leq 1$, that is for any real value $q \geq 2$.
The Ising case ($q=2$) corresponds to a pure annihilation problem 
($p_c=0$ and $p_a=1$), whereas the $q \to \infty$ limit corresponds to
a pure coagulation case ($p_c=1$ and $p_a=0$).
It turns out that the later case is much simpler to study
than any finite $q$ case. In particular, in the $q=\infty$ limit,
simple random walk arguments
are sufficient to obtain the distribution of domain sizes \cite{DGY91} \cite{DZ96}
and the exponent for persistent spins \cite{DBG94},
whereas the computation of the corresponding quantities for any finite $q$ is
much more involved \cite{DHP1,DHP2,DZ96}.
The relative simplicity of coagulation models
($A+A\longrightarrow A$), with possibly the back reaction ($A \longrightarrow A+A$)
or a random input of $A$-particles, or localized sources of $A$-particles, is in fact related to the possibility to write closed kinetic
equations for the ``one-empty-interval probabilities", i.e.
the probabilities that a given interval contains no A-particle \cite{DBA88,DBA89,BDBA89,DB90,BABD90,Doe92,BA95,KPWH95,DHP1}. 
This approach may be generalized to write closed kinetic equations for
 the probabilities to have many disconnected empty regions \cite{Doe92,DHP1}, but all these many-empty-interval probabilities 
may in fact be expressed in terms of the one-empty-interval probabilities
alone \cite{DHP1}. This means that all the information on the coagulation model
is actually contained in these one-empty-interval probabilities. 

The method that has been followed to study the general $q$ case
\cite{Der95,DHP1,DHP2,DZ96} has been to relate
 the zero-temperature Glauber dynamics 
of the $q$-state Potts model to another reaction-diffusion problem which
is a pure coagulation problem ($A+A \longrightarrow A$) for any $q$,
in contrast with the reaction-diffusion model of domain-walls $W$ (\ref{Wrules}).
A simple way to implement the zero-temperature Glauber dynamics 
consists in updating the spins according to
\begin{eqnarray}
 S_i(t) =  & S_{i-1}(t-dt) \qquad &\hbox{with probability} \ \ dt \\
 S_i(t) =  & S_{i+1}(t-dt) \qquad &\hbox{with probability} \ \ dt \\
 S_i(t) =  & S_{i}(t-dt)   \qquad &\hbox{with probability} \ \ (1-2dt) 
\label{update}
\end{eqnarray}
Tracing back in time the sequence of spins
responsible for the value $S_i(t)$ of spin $i$ at time $t$ therefore defines a random walk going
backwards in time and leading to some spin $S_j(0)$ of the initial condition.
If we are interested in the values $S_i(t)$ and $S_i(t')$
of the same spin at two different times \cite{Der95} \cite{DHP1} \cite{DHP2},
we have to consider the corresponding two random walks 
starting respectively at time $t$ and $t'$ from site $i$ and going backwards
in time, and study whether they merge at some point before the initial time $t=0$ in which case $S_i(t)=S_i(t')$, 
or whether they do not meet up to time $t$, in which case $S_i(t)=S_i(t')$
with probability ${1 \over q}$ due to the randomness of the initial condition.
This is the starting point for studying the probability that a given spin $i$
does not flip up to time $t$ \cite{DHP2}.
The same type of reasoning can be applied to compare the values $S_i(t)$ and $S_j(t)$ of two different spins at time $t$ to lead to the distribution
of domain sizes \cite{DZ96}.
In this approach, the Glauber dynamics of the $q$-state Potts model
is thus formulated in terms of random walks going backwards in time
that merge whenever they meet ($A+A \longrightarrow A$), and the parameter $q$
only appears through the property that two different sites 
of the random initial condition have the same value with probability ${1 \over q}$. The problem has now therefore a mathematical meaning for any $q \geq 1$,
in contrast with the initial reaction-diffusion model of domain-walls $W$ (\ref{Wrules})
defined for $q \geq 2$ only.
This model can moreover be given a physical meaning for any real value $q \geq 1$ if one considers the Ising case with a random initial condition presenting a
non-zero magnetization  $m \in [-1,+1]$, in which each spin has initially the value $(+)$ with probability $p_+={{1+m}\over 2}$ and the value $(-)$ with probability $p_-={{1-m}\over 2}$ \cite{SM95a,SM95b,BNFK96,DZ96}. Indeed the probability
that a (+) spin does not flip up to time $t$ or the distribution of (+)
domain sizes will be given by the corresponding results for the previous
$q$-state Potts model with the correspondence
\be
{1\over q}=p_+={{1+m}\over 2}
\label{q-m}
\ee

This article is devoted to the following generalization of the problem 
of the exponent of persistent spins for the zero-temperature Glauber dynamics 
of the $q$-state Potts model.
We consider a B-particle that diffuses with a diffusion constant $D_B = \lambda
D_W$ that is different from the diffusion constant $D_W$ of domain walls 
($\lambda \in [0,+\infty)$),
and that disappears whenever it meets a domain wall $W$
\[
B+W  \longrightarrow \emptyset \qquad \hbox{with probability 1}
\]
The question is : what is the exponent $\theta(q,\lambda)$ that describes
the survival probability of the B-particle
\be
P_B(t,q,\lambda) \oppropto_{t \to \infty} t^{- \theta(q,\lambda)}
\ee
as a function of $q \geq 1$ and $\lambda \geq 0$ ?
The exponent is expected to depend explicitly on the ratio $\lambda={D_B \over D_W}$
because of the interplay between the diffusion of $B$
and the domain coarsening of
the kinetic Potts model. 
From the point of view of reaction-diffusion models,
the problem considered here is ``an impurity problem" \cite{KBNR94,H95}, in which there
is a small concentration of $B$-particles in an homogeneous background of $W$-particles,
so that one can neglect reactions among impurities and the influence of impurities-background reactions on the background properties. In this language, the problem of the fraction of persistent spins in the kinetic Potts model can be reformulated as
a ``static impurity problem" \cite{KBNR94}.

What do we know about the exponent $\theta(q,\lambda)$?
It is clear from the definition of the model that $\theta(q,\lambda)$
is an increasing function of $q$ at fixed $\lambda$, 
and an increasing function of $\lambda$ at fixed $q$. 
For $q=1$ there is no domain walls $W$ so that 
the exponent vanishes in this limit $\theta(q=1, \lambda)=0$.
 For a fixed B-particle ($\lambda=0$), the survival probability of the B-particle
reduces to the
probability that a given spin is not crossed by any domain wall up to time $t$, which is also the probability that a given spin  
does not flip up to time $t$, and the exact expression recently obtained
for this exponent reads \cite{DHP1} \cite{DHP2}
\be
\theta(q,\lambda=0) =-{1\over 8} +{2\over \pi^2} \arccos^2\left({{2-q} \over {{\sqrt 2} q}}\right)  
\label{lambda0}
\ee
in agreement with previous numerical results \cite{DBG94,Sta94,Der95}. 
The exponent $\theta(q,\lambda)$ is also known in the particular case
 $q=2$ and $\lambda=1$ where the $B$-particle can be considered
as a domain wall $W$, and where the dynamics of domain walls reduces to a pure annihilation model ($W+W \longrightarrow \emptyset$) \cite{TMC83,TW83,Lus86}
\be
\theta(q=2,\lambda=1) ={1\over 2}
\label{theta21}
\ee
We have not been able to get an exact expression of the exponent
$\theta(q,\lambda)$ in the general case $q>1$ and $\lambda>0$,
but we have obtained various asymptotic behaviors.
The paper is organized as follows.
In Section \ref{sec-qinfty} we recall how to derive the value of the exponent $\theta(q,\lambda)$
in the particularly simple case $q=\infty$ 
 and arbitrary $\lambda$ \cite{BA88,KBNR94} 
\be
\theta(q=\infty,\lambda)=
{\pi \over {2   \arccos \left({\lambda \over {1+\lambda}}\right) }} 
\ee
 In section \ref{sec-qepsilon}, we extend the approach described in 
reference \cite{DHP1} by Derrida, Hakim and Pasquier to obtain the first correction in $\epsilon=q-1$ of the exponent
for any $\lambda$
\be
\theta(q=1+\epsilon,\lambda) \equiv \epsilon  \left({ {\sqrt{2\lambda+1}} \over
 {\pi -\arccos \left({\lambda \over {1+\lambda}}\right)}} \right)+o(\epsilon)
\ee
As previously explained, this exponent characterizes the decay of the probability that a B-particle remains in a (+) domain up to time $t$
for the Glauber dynamics of the Ising model starting from a random
initial condition characterized by a strong magnetization $m=1-2\epsilon$ (\ref{q-m}).
In section \ref{lambdapetit}, we generalize the approach developed in 
reference  \cite{DHP2} by Derrida, Hakim and Pasquier to study the first-order perturbation in $\lambda$ around the result (\ref{lambda0})
\be
\theta (q, \lambda)=\theta (q, 0)+\lambda \phi(q) +o(\lambda)
\ee
 but the expression obtained for $\phi(q)$ is unfortunately
quite complicated (see eq (\ref{phiqfinal}) below).

To make the reading easier, we gather here the useful notations used in the various parts of the paper
\begin{eqnarray*}
&\alpha \equiv \alpha(\lambda) &\equiv \arccos\left({\lambda \over {1+\lambda}}\right)
\\
& \nu \equiv \nu(\lambda) &\equiv {1 \over 2} \left(1+{\alpha \over \pi} \right)
\\
 &\mu \equiv \mu(q) &\equiv {{q-1} \over q^2}
\\
& \delta \equiv \delta(q)  &\equiv {1 \over \pi} \arccos \left(-4 \mu \right)
\\
&\hat \delta \equiv \hat \delta(q) &\equiv {2 \over \pi} 
\arccos\left({ {2-q} \over {{\sqrt 2 } q}}\right)
\end{eqnarray*}

\section{Direct study of the exponent $\theta(\lowercase{q},\lambda)$
 for $\lowercase{q}=\infty$ and arbitrary $\lambda$}

\label{sec-qinfty}

For $q=\infty$, the dynamics of domain walls (\ref{Wrules})
reduces to a pure coagulation model
\[
W+W \longrightarrow W \qquad \hbox{with probability} \ 1 
\]
The problem of the B-particle survival is therefore reduced to 
a three-body problem, since the two nearest domain walls enclosing $B$
evolve only by diffusion, and cannot disappear when meeting the next nearest  domain walls. Let us introduce
the joint probability $\psi(x_1,x,x_2,t)$ that the B-particle has not yet
disappeared at time $t$ and is 
at position $x$, with the next domain wall to its left being at position $x_1<x$
and the next domain wall to its right being at position $x_2>x$.
This joint probability evolves 
in time according to the diffusion equation 
(where, for simplicity, we have set the diffusion constant of domain walls $D_W$ equal to $1$)
\be
{{\partial \psi} \over {\partial t}} = {{\partial^2 \psi} \over {\partial x_1^2}} 
+ {{\partial^2 \psi} \over {\partial x_2^2}}+\lambda {{\partial^2 \psi} \over {\partial x^2}} \qquad \hbox{for} \qquad x_1<x<x_2  
\label{diffpsi}
\ee
with the absorbing boundary conditions
$\psi(x,x,x_2,t)=0= \psi(x_1,x,x,t)$. 
The survival probability of the B-particle
\be
P_B(t,\infty,\lambda)=\int_{-\infty}^{+\infty}dx \int_{-\infty}^{x}dx_1 \int_{x}^{+\infty}dx_2  
\ \ \psi(x_1,x,x_2,t)
\label{survieB}
\ee
exhibits the asymptotic behavior (see Appendix A) 
\be
P_B(t,q=\infty,\lambda)\oppropto_{t \to \infty} \  t^{-{\pi \over {2\alpha(\lambda)}}}
 \qquad \hbox{where} \qquad
\alpha(\lambda) \equiv \arccos\left({\lambda \over {1+\lambda}}\right)\ee
We therefore recover the value of the exponent (\cite{BA88} \cite{KBNR94} and
references therein)
\be
\theta(q=\infty,\lambda)={\pi \over {2\alpha(\lambda)}}
\label{thetaqinf}
\ee
The angle $\alpha(\lambda)$ decreases from $\alpha(\lambda=0)={\pi \over 2}$
to $\alpha(\lambda=\infty) =0$, so that $\theta(q=\infty,\lambda)$
grows from $\theta(q=\infty,\lambda=0)=1$ to 
$\theta(q=\infty,\lambda=\infty)=\infty$. 

\section{Exponent $\theta(\lowercase{q}=1+\epsilon,\lambda)$ at first order in $\epsilon$
for any $\lambda$}

\label{sec-qepsilon}

In this section, we follow the approach described in 
reference \cite{DHP1} by Derrida, Hakim and Pasquier
and only mention the modifications that have to be made for the case 
we are interested in here.

\subsection{Equivalence with a coagulation model on a large ring}

Finite-size scaling arguments imply that the exponent 
$\theta(q,\lambda)$ also appears in the zero-temperature Glauber 
dynamics of the $q$-state Potts model defined on a ring of finite length $L$.
It describes in this case the decay with the size $L$ of the probability
for the B-particle to survive indefinitely
\be
{\cal P}_B(L,q, \lambda) \sim P_B(t \sim L^2,q,\lambda) \sim L^{-2 \theta(q,\lambda)}
\ee

As explained in \cite{DHP1}, when the spin-values seen by the B-particle are traced back in time, one obtains random walks going backwards in time
(that we will call A-particles from now on),
that merge whenever they meet $(A+A \longrightarrow A)$, and that connect all the spin values seen by the B-particle to various ancestors belonging to the random initial configuration. The probability for $m$
ancestors of the random initial condition to have the same ``color" of the $q$-state Potts model is simply $\left({1 \over q} \right)^{(m-1)}$. As a consequence, the survival probability may be expressed as \cite{DHP1}
\be
{\cal P}_B(L,q, \lambda) = \sum_{m=1}^L {1 \over q^{m-1}} \  p_L(m,\lambda)
\label{dev1/q}
\ee
where $p_L(m,\lambda)$ is the probability of finding $m$ particles on a ring
 of $L$ sites in the steady-state of the following one-species A-particle coagulation problem :
there is a moving ``source" (corresponding to the B-particle) that is always occupied by a A-particle, and
the (L-1) other sites may be either occupied or empty. During each infinitesimal time-step $dt$, each A-particle hops with probability $dt$ to its right neighbor
and with probability $dt$ to its left neighbor, and does not move with probability $(1-2dt)$,
in correspondence with the updating rules of the $T=0$ Glauber dynamics (\ref{update}). If two particles occupy the same site, they instantaneously coagulate $(A+A \longrightarrow A)$. Whenever the A-particle
being on the source jumps to one of its neighbors, a new A-particle is instantaneously produced on the source. The additional rules for the dynamics
of the source are the following :  during each infinitesimal time step $dt$, the source hops with probability $\lambda dt$ to its right neighbor
and with probability $\lambda dt$ to its left neighbor, and does not move with probability $(1-2 \lambda dt)$. Whenever the source moves, the A-particle 
that was occupying the position of the source remains on this site,
and a new A-particle is instantaneously created at the new position of the source.

To compute the expression (\ref{dev1/q}) that involves the probabilities 
$p_L(m,\lambda)$ characterizing the steady state of the coagulation model, 
we introduce generalized ``empty-interval probabilities" 
\cite{DBA88,DBA89,BDBA89,DB90,BABD90,Doe92,BA95,KPWH95,DHP1}. 
We first define the conditional probabilities 
$b_{i,j}^{\{ {\cal S}\}}(t)$ , $(1 \leq i <j\leq L)$, that the segment $\{s(t)+i, s(t)+j-1\}$
contains no A-particle, for a given source trajectory $\{ {\cal S}\}=\{s(\tau), \tau \geq 0\}$
representing a particular realization of the source random walk.
They evolve in time according to
\be \ba
b_{i,j}^{\{ {\cal S}\}}(t+dt)=b_{i,j}^{\{ {\cal S}\}}(t)+
dt \ \delta_{s(t+dt),s(t)} \  \bigg[ b_{i+1,j}^{\{ {\cal S}\}}(t) +
b_{i-1,j}^{\{ {\cal S}\}}(t)+b_{i,j+1}^{\{ {\cal S}\}}(t)
+b_{i,j-1}^{\{ {\cal S}\}}(t) -4 b_{i,j}^{\{ {\cal S}\}}(t) \bigg]
\\   \\
+
dt \ \delta_{s(t+dt),s(t)+1} \ \bigg[ b_{i-1,j-1}^{\{ {\cal S}\}}(t) -
b_{i,j}^{\{ {\cal S}\}}(t) \bigg]
+
dt \ \delta_{s(t+dt),s(t)-1} \  \bigg[ b_{i+1,j+1}^{\{ {\cal S}\}}(t) -
b_{i,j}^{\{ {\cal S}\}}(t) \bigg]
\ea \label{bs}
\ee
and satisfy the boundary conditions
\be
b_{0,j}^{\{ {\cal S}\}}(t)=0=b_{i,L+1}^{\{ {\cal S}\}}(t)
\qquad \hbox{and} \qquad b_{i,i}^{\{ {\cal S}\}}(t)=1
\ee
The average of $b_{i,j}^{\{ {\cal S}\}}(t)$ over the random walk trajectories 
of the source denoted by
\be
B_{i,j}(\lambda, t) = \  <b_{i,j}^{\{ {\cal S}\}}(t)>
\ee
evolve in time according to
\be \ba
{\partial \over \partial t}B_{i,j}(\lambda, t)=B_{i+1,j}(\lambda, t)+B_{i-1,j}(\lambda, t)+B_{i,j+1}(\lambda, t)+B_{i,j-1}(\lambda, t)-4B_{i,j}(\lambda, t)
\\   \\
+\lambda \bigg[B_{i+1,j+1}(\lambda, t)+B_{i-1,j-1}(\lambda, t)-2B_{i,j}(\lambda, t) \bigg]
\ea \ee
and converge at large time towards stationary probabilities $B_{i,j}(\lambda)$
that are solutions of
\be \ba
B_{i+1,j}(\lambda)+B_{i-1,j}(\lambda)+B_{i,j+1}(\lambda)+B_{i,j-1}(\lambda)-4B_{i,j}(\lambda)
\\   \\
+\lambda \bigg[B_{i+1,j+1}(\lambda)+B_{i-1,j-1}(\lambda)-2B_{i,j}(\lambda) \bigg]=0
\ea \label{BS}
\ee
together with the boundary conditions
$B_{0,j}(\lambda)=0=B_{i,L+1}(\lambda) $ and $B_{i,i}(\lambda)=1$

We may also define the conditional probabilities 
$b_{i_1,i_2,\cdots,i_{2n-1},i_{2n}}^{\{ {\cal S}\}}(t)$ , $(1 \leq i_1 <i_2
<\cdots <i_{2n} \leq L)$, that there is no A-particle in any of the segments
$\{s(t)+i_1, s(t)+i_2-1\}, \ldots, \{s(t)+i_{2n-1}, s(t)+i_{2n}-1\}$
 for a given source trajectory $\{ {\cal S}\}=\{s(\tau), \tau \geq 0\}$.
These conditional probabilities satisfy evolution equations analogous to
(\ref{bs}), with obvious boundary conditions for coinciding indices.
We have for example
\be \ba
b_{i,j,k,l}^{\{ {\cal S}\}}(t+dt)=b_{i,j,k,l}^{\{ {\cal S}\}}(t)+
dt \ \delta_{s(t+dt),s(t)} \  \bigg[ b_{i+1,j,k,l}^{\{ {\cal S}\}}(t) +
b_{i-1,j,k,l}^{\{ {\cal S}\}}(t)+b_{i,j+1,k,l}^{\{ {\cal S}\}}(t)
+b_{i,j-1,k,l}^{\{ {\cal S}\}}(t) 
\\  \\ +b_{i,j,k+1,l}^{\{ {\cal S}\}}(t) +b_{i,j,k-1,l}^{\{ {\cal S}\}}(t)
+b_{i,j,k,l+1}^{\{ {\cal S}\}}(t)+b_{i,j,k,l-1}^{\{ {\cal S}\}}(t)
-8 b_{i,j,k,l}^{\{ {\cal S}\}}(t) \bigg]
\\   \\
+
dt \ \delta_{s(t+dt),s(t)+1} \ \bigg[ b_{i-1,j-1,k-1,l-1}^{\{ {\cal S}\}}(t) -
b_{i,j,k,l}^{\{ {\cal S}\}}(t) \bigg]
+
dt \ \delta_{s(t+dt),s(t)-1} \  \bigg[ b_{i+1,j+1,k+1,l+1}^{\{ {\cal S}\}}(t) -
b_{i,j,k,l}^{\{ {\cal S}\}}(t) \bigg]
\label{bss}
\ea \ee

The averages $<b_{i_1,i_2,\cdots,i_{2n-1},i_{2n}}^{\{ {\cal S}\}}(t)>$
over the realizations of the source trajectories converge at large time 
to stationary probabilities $B_{i_1,i_2,\cdots,i_{2n-1},i_{2n}}(\lambda)$
satisfying equations generalizing (\ref{BS}), as for example
\be \ba
B_{i+1,j,k,l}(\lambda)+B_{i-1,j,k,l}(\lambda)+B_{i,j+1,k,l}(\lambda)+B_{i,j-1,k,l}(\lambda)
\\  \\ +B_{i,j,k+1,l}(\lambda)+B_{i,j,k-1,l}(\lambda)
+B_{i,j,k,l+1}(\lambda)+B_{i,j,k,l-1}(\lambda)
\\   \\
-8B_{i,j}(\lambda)
+\lambda \bigg[B_{i+1,j+1,k+1,l+1}(\lambda)+B_{i-1,j-1,k-1,l-1}(\lambda)-2B_{i,j,k,l}(\lambda) \bigg]=0
\ea 
\ee

The generalization of the identity (10) of \cite{DHP1} gives the survival
probability of the B-particle in terms of the whole hierarchy of the
mean empty-interval probabilities $B_{i_1,i_2,\cdots,i_{2n-1},i_{2n}}(\lambda)$ as
\be
{\cal P}_B(L,q, \lambda) = {1 \over {q^{L-1}}} \bigg[1+ 
\sum_{1 \leq i <j \leq L} (q-1)^{j-i} B_{i,j} (\lambda) + 
\sum_{1 \leq i <j <k<l \leq L} (q-1)^{j-i+l-k} B_{i,j,k,l} (\lambda)
+ \ldots \bigg]
\label{devBij}
\ee
The key to solve the coagulation model in the case of a fixed source \cite{DHP1}
(corresponding to the $\lambda=0$ case here) was the possibility to write 
the mean many-empty-interval probabilities $B_{i_1,i_2,\cdots,i_{2n-1},i_{2n}}(\lambda=0)$
of arbitrary order as Pfaffians of the mean one-empty-interval probabilities $B_{i,j}(\lambda=0)$ alone, 
as for example
\be
B_{i,j,k,l}(0) \ = \ B_{i,j}(0)B_{k,l}(0)+B_{i,l}(0)B_{j,k}(0)
-B_{i,k}(0)B_{j,l}(0)
\label{Pfaff0}
\ee

For the case of a moving source, all these Pfaffians relations still hold for a given realization
${\{ {\cal S}\}}$ of the source trajectory. It is for example easy to check that
\be
b_{i,j,k,l}^{\{ {\cal S}\}}(t)=b_{i,j}^{\{ {\cal S}\}}(t) \ b_{k,l}^{\{ {\cal S}\}}(t)+b_{i,l}^{\{ {\cal S}\}}(t) \ b_{j,k}^{\{ {\cal S}\}}(t)-b_{i,k}^{\{ {\cal S}\}}(t) \ b_{j,l}^{\{ {\cal S}\}}(t)
\label{Pfaffien}
\ee
since the right hand side and the left hand side
 evolve in time according to the same equation (\ref{bss}) and satisfy the same boundary
conditions.
However, these Pfaffians relations that hold for a given realization
of the source trajectory are no longer valid for the mean probabilities $B_{i_1,i_2,\cdots,i_{2n-1},i_{2n}}(\lambda)$,
because Pfaffians involve products and the mean of a product is of course not 
equal to the product of means. 
In particular, equation (\ref{Pfaff0}) is no longer valid for $\lambda \neq 0$
\be
B_{i,j,k,l}(\lambda) \ \neq \ B_{i,j}(\lambda)B_{k,l}(\lambda)+B_{i,l}(\lambda)B_{j,k}(\lambda)-B_{i,k}(\lambda)B_{j,l}(\lambda)
\label{nPfaffien}
\ee
Equation (\ref{devBij}) giving the survival
probability of the B-particle in terms of the whole hierarchy of the
mean many-empty-interval probabilities $B_{i_1,i_2,\cdots,i_{2n-1},i_{2n}}(\lambda)$
is therefore much more difficult to use for $\lambda\neq 0$.

This is why in the following, we only compute the mean one-empty-interval probabilities $B_{i,j} (\lambda)$ in the limit of a large system $L \to \infty$, and use the following expansion of the survival probability (\ref{devBij}) in $\epsilon=q-1$
\be
{\cal P}_B(L,q=1+\epsilon, \lambda) = 1+\epsilon \ 
\sum_{i=1}^{L-1} \bigg( B_{i,i+1} (\lambda) -1 \bigg) +o(\epsilon^2) 
\label{survepsi}
\ee
to obtain the first order in $\epsilon$ of the exponent $\theta(q=1+\epsilon,\lambda)$.

\subsection{Mean one-empty-interval probabilities $B_{i,j}(\lambda)$ for a large system $L \to \infty$}

\label{BLinfty}

For large $L$, $B_{i,j}(\lambda)$ becomes a continuous function $\beta_{\lambda}(x={i \over L},y={j \over L})$ which satisfies the continuous
version of (\ref{BS})
\be
(1+\lambda) \ \left({{\partial^2 \beta} \over {\partial x^2}}+{{\partial^2 \beta} \over {\partial y^2}}\right)+2\lambda \ {{\partial^2 \beta} \over {\partial x \partial y}}=0
\ee
in the triangle $0 \leq x \leq y \leq 1$ together 
with the boundary conditions
$\beta_{\lambda}(0,y) =0 = \beta_{\lambda}(x,1) 
$ and $\beta_{\lambda}(x,x)=1$.
To eliminate the non-diagonal term, we perform the change of coordinates
\be
X=x \qquad \hbox{and} \qquad Y={{(1+\lambda)y-\lambda x} \over 
{\sqrt{2\lambda +1}}}
\ee
The problem is now reduced to solving the Laplace equation
\be
{{\partial^2 \beta_{\lambda}} \over {\partial X^2}} 
+ {{\partial^2 \beta_{\lambda}} \over {\partial Y^2}} =0   
\label{LT}
\ee
in the triangle of vertices $O(X_O=0,Y_O=0)$, $A \left(X_A=0, 
Y_A={{1+\lambda} \over {\sqrt{2\lambda +1}}}\right)$ 
and $C \left(X_C=1,Y_C={{1} \over {\sqrt{2\lambda +1}}}\right)$,
with angles $\hat{OAC}=\alpha \equiv \arccos({\lambda \over {\lambda+1}})$
and $\hat{ACO}=\hat{COA}=\gamma\equiv {{\pi-\alpha}\over 2}$
(see Figure 1) with the boundary conditions
\be
\beta_{\lambda}(X,Y)=0 \ \  \hbox{on segments OA and AC}
\qquad \hbox{and} \qquad \beta_{\lambda}(X,Y)=1 \ \  \hbox{on segment OC}
\label{CLT}
\ee

A convenient way to solve this problem is to use methods of complex
analysis and conformal transformations.
We introduce the function $\tilde \beta (w)$ of the complex variable $w=u+iv$
\be
\tilde \beta (w) = \Im \bigg[ {1 \over \pi} \ln \left({{w-1} \over w} \right)
\bigg]
\label{tbw}
\ee
where $\Im$ denotes the imaginary part.
This function $\tilde \beta (w)$ satisfies the Laplace equation on the upper half-plane
 $\{v \geq 0\}$
 and the following boundary conditions on the real axis $\{v=0\}$
\be
\tilde \beta \bigg( u \in (-\infty,0),v=0 \bigg)=0 
=\tilde \beta \bigg( u \in (1,+\infty),v=0 \bigg)
\qquad \hbox{and} \qquad 
\tilde \beta \bigg( u \in (0,1),v=0 \bigg)=1
\ee
We now consider the conformal transformation
\be
Z(w)= K \  \int_0^w d\xi {1 \over {\vert \  \xi (1- \xi)\vert \  ^{\nu}}} \  e^{i( {\alpha \over 2}
+ \nu( \pi -\arg(\xi)-\arg(\xi-1)))}
\ee
with the notations 
\be
 \nu={1 \over 2} \left(1+{\alpha \over \pi} \right)
\qquad \hbox {and} \qquad 
 K={ {\Gamma(2-2\nu)} \over {\Gamma^2(1-\nu) \ \cos({\alpha \over 2})}} 
\label{Zw}
\ee
This transformation maps the upper half-plane of the complex plane $w$ into the interior of the triangle $OAC$ described above in the complex plane $Z=X+iY$.
We have more precisely $Z_O=Z(w=0)$, $Z_C=Z(w=1)$ and $Z_A=Z(w=\pm \infty)$.
The function $\beta_{\lambda}(X,Y)$ of (\ref{LT} - \ref{CLT}) therefore
reads
\be
\beta_{\lambda}(Z=X+iY)=\tilde \beta \big[ w(Z) \big]
=\Im \bigg[ {1 \over \pi} \ln \left({{w(Z)-1} \over {w(Z)}} \right)
\bigg]
\ee
where $w(Z)$ is the inverse mapping of $Z(w)$ (\ref{Zw}).
 Unfortunately, the inverse mapping $w(Z)$ cannot be explicitly written
for arbitrary $Z$ in the triangle. However, approximate explicit forms may be 
written locally.

\subsection{Use of $B_{1,L}(\lambda)$ to recover $\theta(q=\infty,\lambda)$  }

The exponent $\theta(q=\infty,\lambda)$ (\ref{thetaqinf})
may be recovered from the asymptotic behavior of $B_{1,L}(\lambda)$
since (\ref{dev1/q}) reduces for $q=\infty$ to
\be
{\cal P}_B(L,q=\infty, \lambda) =   p_L(1,\lambda) 
\ee
and the probability of having only exactly one particle on the ring is simply the probability $B_{1,L}(\lambda)$ of having no particle except on the source.
So finally
\be
{\cal P}_B(L,q=\infty, \lambda) =B_{1,L}(\lambda) \simeq
\beta_{\lambda}(x \sim {1 \over L},y \sim 1-{1 \over L}) \propto L^{- {2\theta(\infty,\lambda)}}
\ee
We thus have to consider the function $ \beta_{\lambda}(Z)$ 
for $Z$ near the vertex A of the triangle.
 In the neighborhood of $Z_A$,
$w(Z=Z_A+r   e^{i(-{\pi \over 2}+\phi)})$ reads approximatively for 
$\phi \in [0,\alpha]$ and small enough $r$
\be
w \bigg(Z=Z_A+r  e^{ i(-{\pi \over 2}+\phi)} \bigg) \opsimeq_{r \to 0} 
\ - \bigg[{\alpha \over \pi} {r \over K} \bigg]^{\pi \over \alpha}
 \   e^{ -i \pi {\phi \over \alpha}}
\ee
and we get
\be
\beta_{\lambda}(Z=Z_A+r  e^{ i(-{\pi \over 2}+\phi)}) 
\opsimeq_{r \to \infty}\Im \bigg[- {1 \over \pi} {1 \over {w(Z=Z_A+r  e^{ i(-{\pi \over 2}+\phi)})}}\bigg]
={1 \over \pi} \bigg[{{\alpha r} \over { \pi K}} \bigg]^{\pi \over \alpha} 
\sin \left(\pi {\phi \over \alpha} \right)
\ee

We finally obtain
\be
{\cal P}_B(L,q=\infty, \lambda) 
 \opsimeq_{L \to \infty} \beta_{\lambda}(x \sim {1 \over L},y \sim 1-{1 \over L}) \sim \beta_{\lambda}(Z \sim Z_A+{1 \over L}  e^{
i(-{\pi \over 2}+\phi)}) 
\propto L^{- {\pi \over \alpha}}
\ee
and so recover again the exponent (\ref{thetaqinf}).

\subsection{ Exponent $\theta(q=1+\epsilon,\lambda)$ at first order in $\epsilon$}

\label{foepsilon}

To obtain the first order in $\epsilon$ of the exponent
\be
\theta(q=1+\epsilon,\lambda)\equiv \epsilon \ a(\lambda)+o(\epsilon)
\ee 
we only have to consider the $B_{i,i+1}(\lambda)$ since $a(\lambda)$ is given by
the leading behavior (\ref{survepsi})
\be
\sum_{i=1}^{L-1} \big(1-B_{i,i+1}(\lambda) \big) \opsimeq_{L \to \infty} 
2 a(\lambda) \ln L 
\label{(q-1)ln}
\ee

We may locally invert $Z(w)$ around $Z_B=Z(0)$
and obtain $w(Z=r  e^{i({\pi \over 2}-\phi)}) $ for $\phi \in [0,\gamma]$ and for small enough $r$
\be
w(Z=r  e^{i({\pi \over 2}-\phi)}) \opsimeq_{r \to 0} - \bigg[(1-\nu) {r \over K} \bigg]^{1 \over {1-\nu}} \  e^{\displaystyle-i {\phi \over {1-\nu}}}
\ee
We therefore get
\be
\beta_{\lambda}(Z=r  e^{i({\pi \over 2}-\phi)}) 
\opsimeq_{r \to 0} \Im \bigg[ {1 \over \pi} \ln \left(
\bigg[{K \over {(1-\nu)r}}\bigg]^{1 \over {1-\nu}}   e^{i {\phi \over {1-\nu}}}
\right) \bigg] = {\phi \over {\pi (1-\nu)}} = {{\arctan({ X \over Y})} \over {\pi (1-\nu)}}
\ee
or more explicitly in terms of the original coordinates $(x,y)$
\be
\beta_{\lambda}(x \ll 1, y \ll 1) \simeq {2 \over {\pi - \alpha}}
 \arctan \left({{x \sin \alpha} \over {y-x \cos \alpha}}\right)
\ee

As explained in \cite{DHP1}, this small corner $0 \leq x < y \leq 1$
where $\beta_{\lambda}(x, y)$ is of the form $f_{\lambda}({x \over y})$
with the scaling function
\be
f_{\lambda}(u)= {2 \over {\pi - \alpha}}
 \arctan \left({{u \sin \alpha} \over {1-u \cos \alpha}}\right)
\label{sff}
\ee
and the symmetric corner $1-y <1-x \ll 1$, are entirely responsible of the leading behavior of (\ref{(q-1)ln}), that is more explicitly
\be
\sum_{i=1}^{L-1} \big(1-B_{i,i+1}(\lambda) \big) \opsimeq_{L \to \infty} 
2 L \int_{1 \over L} dx \ \bigg[ 1- \beta_{\lambda} \left(x,x+{1 \over L}\right) \bigg]
\ee
\be
\simeq 2 L \int_{1 \over L} dx \ \bigg[ 1- f_{\lambda} 
\left(1-{1 \over {Lx}}\right) \bigg] \simeq 2 f_{\lambda}'(1) \ln L
\ee
The correction $a(\lambda)$ (\ref{(q-1)ln}) is therefore given by
\be
a(\lambda)= f_{\lambda}'(1) ={1 \over {(\pi - \alpha) \tan({\alpha \over 2})}}
\ee
so that finally
\be
\theta(q=1+\epsilon,\lambda) \equiv \epsilon \  \left({ {\sqrt{2\lambda+1}} \over
 {\pi -\arccos \left({\lambda \over {1+\lambda}}\right)}} \right) +o(\epsilon)
\label{thetaepsi}
\ee

\section{Exponent $\theta(\lowercase{q},\lambda)$ at first order in $\lambda$
for any $\lowercase{q}$}

\label{lambdapetit}

In this section, we follow the approach of reference \cite{DHP2} by Derrida, Hakim and Pasquier
and only mention the modifications that have to
 be made for the present study.

\subsection{ Adaptation of the formalism introduced in [6]}

Let us introduce the survival probability $p_B^{\{ \cal S\}}(t,q)$ up to time $t$
of the B-particle for a given trajectory ${\cal S}=\{ s(\tau), \tau \geq 0\}$ 
of the B-particle.
We may also define the analog $R_B^{\{ \cal S\}}(t,q)$ of this survival probability
for the semi-infinite-chain geometry \cite{DHP2}. The survival probability
 on the infinite chain reads then as a generalization of Eq (10) of \cite{DHP2}
\be
p_B^{\{ \cal S\}}(t,q)=R_B^{\{+\cal S\}}(t,q) \ R_B^{\{-\cal S\}}(t,q)
\label{survBS}
\ee
where the notation $\{-\cal S\}$ denotes the mirror-trajectory
$\{-s(\tau) , \tau \geq 0 \}$ of the trajectory ${\cal S}=\{s(\tau), \tau \geq 0\}$.
 The generalization of Eqs (29-30-31)  of \cite{DHP2}
is that the survival probability $R_B^{\{ \cal S\}}(t_1,t_2,q)$ between times $t_1$ 
and $t_2$ for a given trajectory ${\cal S}$ may be expressed as
\be
R_B^{\{ \cal S\}}(t_1,t_2,q)=Q^{\{ \cal S\}}(t_1,t_2,q) \  
e^{ {1 \over 2} T^{\{ \cal S\}}(t_1,t_2,q)}
\label{defQT}
\ee
where, using the notation $\mu={{q-1} \over q^2}$
and denoting by $\partial_1$ the derivative with respect to the first variable
of any function of several variables, we have
\be \ba
T^{\{ \cal S\}}(t_1,t_2,q) =- \displaystyle \sum_{n=1}^{\infty} {{(-2\mu)^n} \over n}
 \int_{t_1}^{t_2} d \tau_1 \cdots  \int_{t_1}^{t_2} d \tau_n 
\\   \\
\partial_1 c^{\{ \cal S\}}(\tau_1,\tau_2) \  \partial_1 c^{\{ \cal S\}}(\tau_2,\tau_3)
\ \cdots  \ \partial_1 c^{\{ \cal S\}}(\tau_n,\tau_1)
\ea \label{Tc}
\ee
and
\be
Q^{\{ \cal S\}}(t_1,t_2,q)= \sqrt{1-\mu {\tilde c}^{\{ \cal S\}}(t_2,t_2,q)}
- (q-1)  \sqrt{-\mu {\tilde c}^{\{ \cal S\}}(t_2,t_2,q)}
\ee
with
\be \ba
{\tilde c}^{\{ \cal S\}}(t_2,t_2,q)=\displaystyle\sum_{n=1}^{\infty} (-2\mu)^n
 \int_{t_1}^{t_2} d \tau_1 \cdots  \int_{t_1}^{t_2} d \tau_n 
\\   \\
c^{\{ \cal S\}}(t_2,\tau_1) \ \partial_1 c^{\{ \cal S\}}(\tau_1,\tau_2) \ \partial_1 c^{\{ \cal S\}}(\tau_2,\tau_3)
\ \cdots  \ \partial_1 c^{\{ \cal S\}}(\tau_n,t_2)
\ea \ee
So the fundamental object needed to compute $R_B^{\{ \cal S\}}(t_1,t_2,q)$ is the probability $c^{\{ \cal S\}}(\tau_1,\tau_2)$ (with $0<\tau_1<\tau_2$)
that, for a given trajectory ${\cal S}=\{s(\tau), 0 \leq \tau \leq \tau_2\}$
 of the source, two Brownian walkers going backwards in time and 
starting respectively at $s(\tau_2)$ at time
$u_2=t-\tau_2$ and at $s(\tau_1)$ at time $u_1=t-\tau_1$ do not meet up to time $t$,
with the boundary condition that they are reflected by the 
reversed-time source trajectory $\{ \sigma(u)=s(t-u) \  , t-\tau_2 \leq u \leq t\}$ (see Figure 2).

The method of images gives that $c^{\{ \cal S\}}(\tau_1,\tau_2)$ 
may be written as
\be \ba
c^{\{ \cal S\}}(\tau_1,\tau_2) =\displaystyle
\int_{\sigma(t)=s(0)}^{\infty} \ dx 
 \ \int_x^{\infty}dy \ 
\bigg[ g^{\{ \cal S\}}(x,t \vert \  s(\tau_1), t-\tau_1)
 \ g^{\{ \cal S\}}(y,t \vert \  s(\tau_2), t-\tau_2)
\\   \\
-g^{\{ \cal S\}}(x,t \vert \  s(\tau_2), t-\tau_2) \ 
g^{\{ \cal S\}}(y,t \vert \  s(\tau_1), t-\tau_1) \bigg]
\ea \label{defc}
\ee
in terms of the probability density $g^{\{ \cal S\}}(x,t \vert \  s(\tau), t-\tau)$ 
that, for a given trajectory ${\cal S}$
 of the source, a Brownian walker going backwards in time and 
starting at $s(\tau)=\sigma(t-\tau)$ at time
$(t-\tau)$ is at site $x$ at time $t$,
with the boundary condition that it is reflected by the 
reversed-time source trajectory $\{ \sigma(u)=s(t-u) \}$.
More explicitly, $g^{\{ \cal S\}}(x,t \vert \  \sigma(u),u)$ satisfies the diffusion equation
\be
 {{\partial } \over {\partial t}}g^{\{ \cal S\}}(x,t \vert \  \sigma(u), u) = 
{{\partial^2 } \over {\partial x^2}} g^{\{ \cal S\}}(x,t \vert \  \sigma(u), u)
 \qquad \hbox{for $x> \sigma(t)$ }  
\label{eqgs}
\ee
together with the initial condition
\be
g^{\{ \cal S\}}(x,t \vert \  \sigma(u), u) \operarrow_{t \to u}
 \delta \big[x-\sigma(u) \big]
\label{cigs}
\ee
and the reflection condition at $x= \sigma(t)$ expressed by the conservation of probability
\be
{d \over dt} \int_{\sigma(t)}^{\infty} dx \ 
g^{\{ \cal S\}}(x,t \vert \  \sigma(u),u) =0 ={{d \sigma(t)}\over dt} 
g^{\{ \cal S\}}(\sigma(t),t \vert \  \sigma(u),u) +
 {{\partial g^{\{ \cal S\}}(x,t \vert \  \sigma(u), u)} \over {\partial x}}
\bigg \vert_{x=\sigma(t)}
\label{clgs}
\ee

Unfortunately, we have not been able to write this probability density $g^{\{ \cal S\}}(x,t \vert \  \sigma(u),u)$ in an explicit simple way 
as a functional of the source trajectory ${\{ \cal S\}}$. As a consequence,
 the average over the sources trajectories that is needed to evaluate
the exponent $\theta(q,\lambda)$ through
\be
P_B(t,q,\lambda)=<p_B^{\{ \cal S\}}(t,q)>=<
R_B^{\{+\cal S\}}(t,q) \ R_B^{-\{\cal S\}}(t,q)> \oppropto_{t \to \infty} t^{-\theta(q,\lambda)}
\label{R+R-}
\ee
seems quite difficult to study.

So from now on, we will restrict ourselves to the evaluation of
the B-survival probability at the first order in the diffusion coefficient $\lambda$ of the B-trajectories.

\subsection{ Perturbation theory in the diffusion constant $\lambda$}

The probability density $g^{\{ \cal S\}}(\sigma(t)+z,t \vert \  \sigma(u), u)$
may be seen as the continuous limit of a discretized version
involving $(N-1)$ intermediate times $t_n=u+n{{t-u} \over N}$ 
$(1 \leq n \leq N-1)$
\be \ba
g^{\{ \cal S\}}(\sigma(t)+z,t \vert \  \sigma(u), u) = 
\\   \\
\displaystyle {\lim_{N \to \infty}
\int_0^{\infty} dz_1 \int_0^{\infty} dz_2 \cdots \int_0^{\infty} dz_{N-1}
 \prod_{n=0}^{N-1} g^{\{ \cal S\}}(\sigma(t_{n+1})+z_{n+1},t_{n+1} \vert \  \sigma(t_{n})+z_{n},t_{n}) }
\ea \ee
with the conventions $t_0=u$, $z_0=0$, $t_N=t$ and $z_N=z$.
In the limit of a vanishing time interval $\Delta t =t_{n+1}-t_n={{t-u} \over N} \to 0$, we may approximate the source trajectory between times $t_n$ and $t_{n+1}$ by a line segment of slope $a_n={{
\sigma(t_{n+1})-\sigma(t_n)} \over {\Delta t}}$. 
The probability density $g^{\{ \sigma(u)=\sigma_0+a u \}}(x,t \vert \  x_0,0)$
solution of (\ref{eqgs}-\ref{cigs}-\ref{clgs}) for the particular case 
of a linear trajectory $\sigma(u)=\sigma_0+a u$ reads
\be \ba
g^{\{ \sigma(u)=\sigma_0+a u \}}(x,t \vert \  x_0,0)= 
\displaystyle {1  \over {2\sqrt{\pi t}}} \ \bigg[
  e^{\displaystyle -{ {(x-x_0)^2} \over {4t}}} + e^{\displaystyle a (x_0-\sigma_0)} \  e^{\displaystyle -{ {(x+x_0-2\sigma_0)^2} \over {4t}}}  
\\   \\
\displaystyle +a \int_{-\infty}^{0} d \eta \  e^{\displaystyle a (x_0-\sigma_0-\eta)} \   e^{\displaystyle -{ {(x+x_0-2\sigma_0-\eta)^2} \over {4t}}} \bigg] 
\ea \ee
So we get
\be
g^{\{ \cal S\}}(\sigma(t)+z,t \vert \  \sigma(u), u) = \lim_{N \to \infty}
\int_0^{\infty} dz_1 \int_0^{\infty} dz_2 \cdots \int_0^{\infty} dz_{N-1}
\prod_{n=0}^{N-1} Q_{a_n}(z_{n+1},z_n,\Delta t)
\label{funct}
\ee
where
\be
Q_{a_n}(z_{n+1},z_n,\Delta t) \equiv g^{\{ \sigma(u)=\sigma(t_n)+a_n u \}}
(\sigma(t_{n+1})+z_{n+1},t_{n}+\Delta t \vert \  \sigma(t_{n})+z_{n},t_{n})
\ee
\be \ba
= \displaystyle {{ e^{\displaystyle-{a_n^2 \over 4}\Delta t-{a_n \over 2} (z_{n+1}-z_n)}} \over {2\sqrt{\pi t}}} \displaystyle \bigg[   e^{\displaystyle-{ {(z_{n+1}-z_n)^2} \over {4 \Delta t}}}
+  e^{\displaystyle-{ {(z_{n+1}+z_n)^2} \over {4 \Delta t}}}
\\   \\
\displaystyle +a_n  \int_{-\infty}^{0} d \eta_n \  e^{\displaystyle-{a_n \over 2}\eta_n}
\  e^{\displaystyle-{ {(z_{n+1}+z_n-\eta_n)^2} \over {4 \Delta t}}} \bigg]
\label{Qan}
\ea \ee
We then obtain the following expansion up to second order in $\{s(u)\}$
\be
g^{\{ \cal S\}}(\sigma(t)+z,t \vert \  \sigma(t-\tau), t-\tau) =
g_0(z,\tau) 
+g_1^{\{ \cal S\}}(z,\tau) +g_2^{\{ \cal S\}}(z,\tau) + \cdots
\ee
with the term of order $0$
\be
g_0(z,\tau)={  {  e^{\displaystyle-{ {z^2} \over {4\tau}}}} \over {\sqrt{\pi \tau}}}
\ee
the term of order $1$ in $\{s(u)\}$
\be
g_1^{\{ \cal S\}}(z,\tau) = - { 1 \over \pi} \int_0^{\tau} du \ { {s(u)}
\over {\sqrt{u(\tau-u)}}} \  
{{\partial^2 } \over {\partial z^2}} \bigg(  e^{\displaystyle-{z^2 \over {4u}}} \bigg)
\ee
and the average of the term of order $2$ in $\{\sigma(u)\}$
\be
< g_2^{\{ \cal S\}}(z,\tau) > = \lambda \tau { {\partial^2 } \over {\partial z^2} } \left( { { e^{\displaystyle-{ {z^2} \over {4\tau} }}} \over {\sqrt{\pi \tau} } } \right)
\label{meang2}
\ee
Note that this last average may be obtained directly from
the average $<g^{\{ \cal S\}}(\sigma(t)+z,t \vert \  \sigma(t-\tau), t-\tau)> $
which is simple for arbitrary $\lambda$ (See Appendix B).
We then obtain the corresponding expansion
\be
c^{\{ \cal S\}}(\tau_1,\tau_2) = c_0(\tau_1,\tau_2)
+c_1^{\{ \cal S\}}(\tau_1,\tau_2)+c_2^{\{ \cal S\}}(\tau_1,\tau_2)+ \cdots
\ee
 of $c^{\{ \cal S\}}(\tau_1,\tau_2)$
defined in (\ref{defc})
\be \ba
c^{\{ \cal S\}}(\tau_1,\tau_2)=  \displaystyle{\int_{0}^{\infty}dz 
\int_z^{\infty}dz'} \ 
\bigg[ g^{\{ \cal S\}}(\sigma(t)+z,t \vert \  \sigma(t-\tau_1), t-\tau_1)
\  g^{\{ \cal S\}}(\sigma(t)+z',t \vert \  \sigma(t-\tau_2), t-\tau_2)
\\   \\
-g^{\{ \cal S\}}(\sigma(t)+z,t \vert \  \sigma(t-\tau_2), t-\tau_2)
\ g^{\{ \cal S\}}(\sigma(t)+z',t \vert \  \sigma(t-\tau_1), t-\tau_1) \bigg]
\ea \ee
with the term of order $0$
\be
 c_0(\tau_1,\tau_2)=1- {4 \over \pi} \arctan {\sqrt{\tau_1 \over \tau_2}}
\label{c0}
\ee
the term of order $1$ in $\{s(u)\}$
\be
c_1^{\{ \cal S\}}(\tau_1,\tau_2)={2 \over {\pi^{3 \over 2}}} \bigg[ 
\int_0^{\tau_1} du \ { {s(u)}
\over { \tau_2+u}} \ \sqrt{ {\tau_2} \over {u(\tau_1-u)} } 
- \int_0^{\tau_2} dv \ { {s(v)}
\over { \tau_1+v}} \ \sqrt{ {\tau_1} \over {v(\tau_2-v)} } \ \ \bigg]
 \label{c1}
\ee
and finally the average of the term of order $2$ in $\{s(u)\}$
\be
<c_2^{\{ \cal S\}}(\tau_1,\tau_2)> ={{2\lambda} \over {\pi^2}} 
\int_0^{\tau_1 } {{du} \over {\sqrt{u(\tau_1 -u)}}} 
\int_0^{\tau_2 } {{dv} \over {\sqrt{v(\tau_2 -v)}}} 
\ \min (u,v) {{v-u} \over  {(v+u)^2}}
\label{c2}
\ee
 To compute the first correction in $\lambda$ of the exponent
\be
\theta (q, \lambda)=\theta (q, 0)+\lambda \phi(q) +o(\lambda)
\label{phiq}
\ee
we have to study the survival of the B-particle between times $t_1$ and $t_2$ (\ref{R+R-}-\ref{defQT})
\be \ba
P_B(t_1,t_2,q,\lambda)= <Q^{\{+ \cal S\}}(t_1,t_2,q) \ Q^{\{- \cal S\}}(t_1,t_2,q) \  e^{{1 \over 2}\left( T^{\{ +\cal S\}}(t_1,t_2,q)+ 
T^{\{ -\cal S\}}(t_1,t_2,q) \right)} >
\\  \\
\displaystyle \oppropto_{t2 \to \infty} t_2^{-\theta(q,\lambda)}
\ea \ee
For $1<q<2$, the prefactor $Q^{\{ \cal S\}}(t_1,t_2,q)\ Q^{\{- \cal S\}}(t_1,t_2,q)$ remains finite in the limit $t_2 \to \infty$ (see \cite{DHP2}
for the detailed study of the $\lambda=0$ case).
Using the expansion
\be
{1 \over 2}\left( T^{\{+ \cal S\}}(t_1,t_2,q)+ 
T^{\{ -\cal S\}}(t_1,t_2,q) \right)=  T_0(t_1,t_2,q)+
{1 \over 2}\left( T_2^{\{+ \cal S\}}(t_1,t_2,q)+ 
T_2^{\{ -\cal S\}}(t_1,t_2,q) \right) + \cdots
\ee 
we find that the correction $\phi(q)$ (\ref{phiq}) is given by the leading behavior of the average
\be
<T_2^{\{+ \cal S\}}(t_1,t_2,q) > \opsimeq_{t_2 \to \infty} - \lambda \phi(q)
\ln \left({t_2 \over t_1} \right)
\ee
This average at first order in $\lambda$ of $ T^{\{+ \cal S\}}(t_1,t_2,q)$ defined in (\ref{Tc}) decomposes into two contributions
\be
<T_2^{\{+ \cal S\}}(t_1,t_2,q) > = \lambda \bigg( {\cal T}_{2}(t_1,t_2,q)
+ {\cal T}_{1,1}(t_1,t_2,q) \bigg)
\ee
with
\be
 {\cal T}_{2}(t_1,t_2,q)=-\sum_{n=1}^{\infty} {{(-2\mu)^n}}
 \int_{t_1}^{t_2} d \tau_1 \cdots  \int_{t_1}^{t_2} d \tau_n 
 {1 \over \lambda} <\partial_1  c_2^{\{ \cal S\}}(\tau_1,\tau_2)>  
\prod_{i=2}^n \partial_1 c_0(\tau_i,\tau_{i+1})
\ee
and 
\be \ba
 {\cal T}_{1,1}(t_1,t_2,q)=- \displaystyle \sum_{n=2}^{\infty} {{(-2\mu)^n}\over 2} \sum_{l=2}^n
 \int_{t_1}^{t_2} d \tau_1 \cdots  \int_{t_1}^{t_2} d \tau_n 
\\   \\
 {1 \over \lambda} <  \partial_1 c_1^{\{ \cal S\}}(\tau_1,\tau_2)
\ \ \partial_1 c_1^{\{ \cal S\}}(\tau_l,\tau_{l+1})>  
\displaystyle \prod_{i \neq 1,l} \partial_1 c_0(\tau_i,\tau_{i+1})
\ea \ee
so that the correction $\phi(q)$ in (\ref{phiq}) will be obtained through
 the asymptotic behavior
\be
 {\cal T}_{2}(t_1,t_2,q)
+ {\cal T}_{1,1}(t_1,t_2,q) \opsimeq_{t_2 \to \infty} -  \phi(q)
\ln \left({t_2 \over t_1} \right)
\label{asphi}
\ee

The asymptotic behaviors of ${\cal T}_{2}(t_1,t_2,q)$
and ${\cal T}_{1}(t_1,t_2,q)$ are studied respectively in Appendix C and Appendix D
and we only give here the final result for $\phi(q)$ obtained in Appendix E
in terms of the auxiliary variable
\be
\hat \delta(q)={2 \over \pi} 
\arccos\left({ {2-q} \over {{\sqrt 2 } q}  } \right)
\ee
which varies in the interval 
$\hat \delta(q=1)= 1/2 <\hat \delta <\hat \delta(q=\infty)=3/2$
\be \ba
\displaystyle \phi(q)=
  {{ \cos (\pi \hat \delta)} \over {\pi^2 \sin (\pi \hat \delta)}}
\left(1+{3 } \hat \delta+{\hat \delta} \Psi\left({1\over 4}\right)
+{1 \over 2} \left[ \left({1 \over 4}-\hat \delta\right) \Psi\left({9 \over 4}-\hat \delta\right) -\left({1 \over 4}+\hat \delta\right) \Psi\left({1 \over 4}+\hat \delta\right) \right]
 \right)
\\  \\
\displaystyle +  {1\over {\pi^2}}
\int_0^1  {dz \over {\sqrt z}} 
\left[ H_+(z,\hat \delta) H_-(z,\hat \delta)
+2 z^{(1-\hat \delta)} 
\left( {{\Gamma(2-\hat \delta)} \over {\Gamma({3 \over 2}-\hat \delta)}} H_+(z, \hat\delta)
-  \left(\hat \delta-{1 \over 2} \right)
 {{\Gamma(\hat \delta)} \over {\Gamma({1 \over 2}+\hat \delta)}} H_-(z, \hat \delta)
\right) \right] 
\label{phiqfinal}\ea \ee
where $\Psi(x) \equiv { {\Gamma'(x)}\over {\Gamma(x)}}$
is the logarithmic derivative of the Gamma function, and where
the functions $H_+(z, \hat \delta)$ and $H_-(z, \hat \delta)$ 
are defined as the series
\be 
H_+(z, \hat \delta)=
{{2\cos(\pi \hat\delta)} \over { \sin(\pi \hat\delta)}} \left[
\sum_{n=1}^{\infty}
 z^{(2n+1-\hat\delta)} \ {{\Gamma \left(2n+{3 \over 2}-\hat\delta\right)} 
\over {\Gamma \left(2n+1-\hat\delta\right)}}
-\sum_{n=0}^{\infty}
 z^{(2n+1+\hat\delta)} \ {{\Gamma \left(2n+{3 \over 2}+\hat\delta\right)} 
\over {\Gamma \left(2n+1+\hat\delta\right)}} \right]
 \ee 
\be \ba
\displaystyle  H_-(z, \hat \delta)
=2  \bigg[ \sum_{n=1}^{\infty}
 z^{(2n+1-\hat\delta)} \ {{\Gamma \left(2n+2-\hat\delta\right)} 
\over {\Gamma \left(2n+{3 \over 2}-\hat\delta\right)}}
+\sum_{n=0}^{\infty}
 z^{(2n+1+\hat\delta)} \ {{\Gamma \left(2n+2+\hat\delta\right)} 
\over {\Gamma \left(2n+{3 \over 2}+\hat\delta\right)}}
\\  \\
\displaystyle   - \sum_{l=0}^{\infty} z^{ \left (l+{1 \over 2}\right)} \  
{{\Gamma \left(l+{3 \over 2} \right)} \over {l!}} \bigg) \bigg]
\ea  \ee 
The function $\phi(q)$ is plotted on Figure 3 in terms of 
the auxiliary variable
$\hat \delta(q)={2 \over \pi} 
\arccos\left({ {2-q} \over {{\sqrt 2 } q}  } \right)$
in the interval 
$\hat \delta(q=1)= 1/2 <\hat \delta <\hat \delta(q=\infty)=3/2$.

It is easy to check that
\be
\phi(q=\infty)={2 \over \pi}
\ee
and
\be
\phi(q=1+\epsilon)=\left({2 \over \pi}-{4 \over \pi^2}\right) \epsilon +o(\epsilon)
\ee
which are consistent with the previous results  (\ref{thetaqinf}) and (\ref{thetaepsi}) 
at first order in $\lambda$.

\vskip 2 true cm

\section{Conclusion}

We have studied in this paper a two-species reaction diffusion system
in the limit where the minority B-species has a very low concentration in comparison with the majority W-species, with the following 
 two-particle reactions : two W-particles either coagulate ($W+W \longrightarrow W$) or annihilate ($W+W \longrightarrow \emptyset$) with the respective probabilities $ p_c=(q-2)/(q-1) $ and $p_a=1/(q-1)$; a B-particle and a W-particle annihilate ($W+B \longrightarrow \emptyset$) with probability $1$. 
We have seen why the exponent $\theta\left(q,\lambda=D_B/D_W\right)$
describing the asymptotic time decay of the minority B-species concentration
could be viewed as a generalization of the exponent of persistent spins
in the zero-temperature Glauber dynamics of the 1D Potts model
starting from a random initial condition.
We have extended the methods introduced by Derrida, Hakim and Pasquier 
for the problem of persistent spins \cite{DHP1} \cite{DHP2} 
to compute the exponent $\theta(q,\lambda)$ 
in perturbation at first order in $(q-1)$ for arbitrary $\lambda$ 
and at first order in $\lambda$ for arbitrary $q$.
Let us now briefly outline the problems that have to be
overcome to go beyond the first order perturbation theories presented here.

In the approach of section \ref{sec-qepsilon},
the survival probability of the B-particle is given in terms of the whole hierarchy of the
mean empty-interval probabilities $B_{i_1,i_2,\cdots,i_{2n-1},i_{2n}}(\lambda)$
of some one-species coagulation model with a randomly moving source (\ref{devBij}).
Here we only computed the mean one-empty-interval probabilities
 $B_{i,j}(\lambda)$, and we were thus limited to the first order in $\epsilon=q-1$.
However, we have seen that for a given realization of the source trajectory,
many-empty-interval probabilities of arbitrary order could be written as Pfaffians of the
one-empty-interval probabilities $b_{i,j}^{\{ {\cal S}\}}(t)$ alone (\ref{Pfaffien}).
One could therefore think of writing the survival probability of the B-particle  
for a given realization of the source trajectory as the square root of some determinant,
as in formula (11) of reference \cite{DHP1}. The remaining problem then consists in
evaluating the average of the square root of this determinant 
over the realizations of the source trajectories.

In the approach of section \ref{lambdapetit},
the fundamental object involved in the expression of the survival probability of the B-particle  
for a given realization of the source trajectory (\ref{survBS}-\ref{defc})
is the probability density $g^{\{ \cal S\}}(x,t \vert \ \sigma(t-\tau), t-\tau)$ 
that, for a given trajectory ${\cal S}$
 of the source, a Brownian walker going backwards in time and 
starting at $\sigma(t-\tau)$ at time
$(t-\tau)$ is at site $x$ at time $t$,
with the boundary condition that it is reflected by the 
source trajectory $\{ \sigma(u) \}$ (\ref{eqgs}-\ref{clgs}).
Here we only computed this probability density $g^{\{ \cal S\}}$ 
in perturbation up to second order in the source trajectory,
and we thus had to restrict ourselves to the first order in the diffusion coefficient $\lambda$. 
However we have seen that the probability density 
$g^{\{ \cal S\}}(x,t \vert \ \sigma(t-\tau), t-\tau)$
could be written as some functional of the source trajectory ${\{ \cal S\}}$ (\ref{funct}-\ref{Qan})
and one could perhaps hope to put this functional in a sufficiently simple form to be able to evaluate the exponent $\theta(q,\lambda)$.

\vskip 2 true cm

\acknowledgments 

I warmly thank Bernard Derrida for having suggested to me the study of the exponent
$\theta(q,\lambda)$ with the methods he had introduced in \cite{DHP1} and \cite{DHP2}
in collaboration with Vincent Hakim and Vincent Pasquier, 
and for his helpful advice I have benefited from during 
the various stages of this work. 
Moreover I am most grateful to him for his numerical computation of the exponent
$\theta(q,\lambda)$ with the finite size scaling method he had previously used in \cite{Der95}, which has allowed me to test the analytical results presented in this paper (in particular (\ref{phiqfinal})). 

I also wish to thank Emmanuel Guitter for a critical reading of the manuscript.

\vfill \eject

\appendix

\section{ Asymptotic behavior of $P_B(\lowercase{t,q}=\infty,\lambda)$}

To obtain the survival probability $P_B(t,q=\infty,\lambda)$ (\ref{survieB}),
it is convenient to introduce the relative coordinates $y_1$ and $y_2$ together with the center of mass $g$ for the diffusion problem~(\ref{diffpsi})
\be
y_1 \equiv x-x_1 >0 \qquad ; \qquad y_2 \equiv x_2-x >0 \qquad ; \qquad g \equiv
{{\lambda(x_1+x_2)+x} \over {2\lambda +1}}
  \ee
The survival probability then reads
\be
P_B(t,\infty,\lambda)=\int_{0}^{+\infty}dy_1  \int_{0}^{+\infty}dy_2  
\ \ f(y_1,y_2,t)
\ee
where the function $f$ satisfies
\be
{1 \over {1+\lambda}}\ {{\partial f} \over {\partial t}} = 
{{\partial^2 f} \over {\partial y_1^2}} 
+ {{\partial^2 \psi} \over {\partial y_2^2}}-{2\lambda \over  {1+\lambda}} \
{{\partial^2 f} \over {\partial y_1 \partial y_2}} \qquad \hbox{for $y_1>0$ and $y_2>0$}  
\ee
with the boundary conditions $f(0,y_2,t)=0=f(y_1,0,t)$.
To eliminate the non-diagonal term, we set
\be
X=y_1 \qquad \hbox{and} \qquad Y={{\lambda y_1+(1+\lambda)y_2} \over 
{\sqrt{2\lambda +1}}}
\ee
The problem is now reduced to the diffusion equation
\be
{1 \over {1+\lambda}}\ {{\partial f} \over {\partial t}} = 
{{\partial^2 f} \over {\partial X^2}} 
+ {{\partial^2 f} \over {\partial Y^2}} \qquad \hbox{in the corner} \ \ 
 Y>{\lambda \over {\sqrt{2\lambda +1}}}\ X >0  
\ee
So finally in polar coordinates, we get
\be
P_B(t,\infty,\lambda)=\int_{0}^{+\infty} \rho \ d\rho  \int_{0}^{\alpha}d\phi  
\ \  g(\rho,\phi,t)
\ee
where $g(\rho,\phi,t)$ is the solution of the diffusion equation
\be
{1 \over {1+\lambda}}{{\partial g} \over {\partial t}} = 
{1\over \rho} {\partial \over \partial \rho } \left(\rho
 {{\partial \psi} \over {\partial \rho}} \right) 
+ {1\over \rho^2} {{\partial^2 \psi} \over {\partial \phi^2}} \qquad \hbox{in the corner
 $0<\phi < \alpha\equiv \arccos\left({\lambda \over {1+\lambda}}\right)$}  
\ee
with the absorbing boundary conditions
$g(\rho,0,t)=0= g(\rho,\alpha,t)$.
The solution $g(\rho,\phi,t)$ for a given initial condition
\be
g(\rho,\phi,t) \operarrow_{t \to 0} \ {1 \over \rho_0} \ \delta(\rho-\rho_0) \ \delta(\phi-\phi_0)
\ee
may be expanded onto an eigenfunction basis as
\be
g(\rho,\phi,t) =  \sum_{m=1}^{\infty} \int_0^{\infty} k \ dk  \
h_{k,m}(\rho,\phi) \  h_{k,m}(\rho_0,\phi_0) \  e^{\displaystyle-(1+\lambda)k^2 t}
\ee
with the eigenfunctions
\be
h_{k,m}(\rho,\phi) = { \sqrt{2 \over \alpha}} \ J_{m{\pi \over \alpha}} (k \rho) \sin \left(m \pi {\phi \over \alpha} \right)
\ee
where $J_{\nu}(z)$ denotes the Bessel function of index $\nu$.
The survival probability of the B-particle may finally be written as
\be
P_B(t,\infty,\lambda)=\sum_{m \ odd} {4 \over {\pi m}} \sin \left(m \pi {\phi_0 \over \alpha} \right) \int_0^{\infty} r \ dr  \int_0^{\infty} q \ dq \ 
  e^{\displaystyle-(1+\lambda)q^2 } \  J_{m{\pi \over \alpha}} (qr) \ 
J_{m{\pi \over \alpha}} \left( {{q \rho_0} \over {\sqrt t}} \right)
\ee
The asymptotic behavior of Bessel function at small argument,
 $ J_{\nu} (z) \simeq {1 \over \Gamma(1+\nu)} \left({z/2}\right)^{\nu} $ as $z \to 0$,
gives the long-time behavior 
\be
P_B(t,q=\infty,\lambda) \opsimeq_{t \to \infty} {4 \over {\pi}} \sin \left( \pi {\phi_0 \over \alpha} \right) { { \Gamma \left (1+{\pi \over {2 \alpha}} \right)}
\over { \Gamma \left (1+{\pi \over { \alpha}} \right)} }
 \left( {{ \rho_0} \over {2\sqrt {(1+\lambda)t}}} \right)^{\pi \over \alpha} 
\oppropto_{t \to \infty} \  t^{-{\pi \over {2\alpha(\lambda)}}}
\ee

\vskip 2 true cm

\section{ Direct evaluation of ${\cal P}_{\lambda}\lowercase{(z,t)=
<g^{\{ \cal S\}}(\sigma(t)+z,t \vert \  \sigma(0), 0)> }$}

The mean probability density $R_{\lambda}(\sigma,x,t )$ 
that the source starting at $\sigma_0=0$ at time $t=0$
is at position $\sigma$ at time t and that the random walker
 emitted by the source at time $t=0$ is at position $x> \sigma$ at time $t$
 is solution of the diffusion equation
\be
 {{\partial } \over {\partial t}} R_{\lambda}(\sigma,x,t )= 
\left( \lambda {{\partial^2 } \over {\partial \sigma^2}} +{{\partial^2 } \over {\partial x^2}} \right)  R_{\lambda}(\sigma,x,t )
 \qquad \hbox{for $x> \sigma$ }  
\ee
together with the initial condition
\be
 R_{\lambda}(\sigma,x,t )\operarrow_{t \to 0}
 \delta \big(\sigma  \big)
\delta \big(x \big)
\ee
and the reflection boundary condition at $x= \sigma$
\be
\left[ (1-\lambda) {{\partial  R_{\lambda}(\sigma,x,t )} \over {\partial x}} -2\lambda {{\partial  R_{\lambda}(\sigma,x,t )} \over {\partial \sigma}}
\right]  \bigg \vert_{x=\sigma} =0
\ee
This boundary condition which may be obtained from the continuous limit
of the discrete-space dynamics, ensures that the partial law
 for the position of the
source is a free Brownian motion of coefficient $\lambda$ as it should
\be
\int_{\sigma}^{\infty} dx \  R_{\lambda}(\sigma,x,t ) =
{{ e^{\displaystyle-{\sigma^2 \over {4 \lambda t}}} \over {2\sqrt{\pi \lambda t}}}}
\ee  
as can be checked by taking the time-derivative of both sides.

The mean probability density  ${\cal P}_{\lambda}(z,t)=
<g^{\{ \cal S\}}(\sigma(t)+z,t \vert \  \sigma(0), 0) >$
of the relative coordinate $z=x-\sigma$ therefore satisfies
the diffusion equation
\be
 {{\partial  {\cal P}_{\lambda}(z,t)} \over {\partial t}}= 
{{\partial^2 {\cal P}_{\lambda}(z,t)} \over {\partial x^2}}
 \qquad \hbox{for $z>0$ }  
\ee
together with the initial condition
\be
{\cal P}_{\lambda}(z,t)\operarrow_{t \to 0}
 \delta \big(z \big)
\ee
and the following boundary condition at $z= 0$
\be
{{\partial {\cal P}_{\lambda}(z,t} \over {\partial z}} \bigg \vert_{z=0} =0
\ee
So it simply reads 
\be
{\cal P}_{\lambda}(z,t) ={  { \displaystyle  e^{\displaystyle - {z^2 \over {4 (1+\lambda) t} } }  }  \over {\sqrt{\pi (1+\lambda) t}}}
\ee
from which we directly recover equation (\ref{meang2}).

\vskip 2 true cm

\section{ Asymptotic behavior of ${\cal T}_{2}\lowercase{(t_1,t_2,q)}$}

To compute ${\cal T}_{2}(t_1,t_2,q)$, we use the notations
\be
c_0(\tau_i,\tau_{i+1}) \equiv g_0\left({\tau_i \over \tau_{i+1}} \right)
\qquad \hbox{and} \qquad 
 < c_2^{\{ \cal S\}}(\tau_1,\tau_2)> \equiv 
\lambda \ g_2\left({\tau_1 \over \tau_2} \right)
\ee
with (\ref{c0})
\be
g_0(z)=1- {4 \over \pi} \arctan {\sqrt{z}}
\ee
and (\ref{c2})
\be
g_2(z)={2 \over {\pi^2}} \int_0^1 
{{dx} \over {\sqrt{x(1-x)}}} 
\int_0^{1}  {{dy} \over {\sqrt{y(1-y)}}} 
\  {\min (zx,y)}  \ {{y-zx} \over {(y+zx)^2}}
\label{g2z}
\ee
and rewrite
\be
 {\cal T}_{2}(t_1,t_2,q)=-\sum_{n=1}^{\infty} {{(-2\mu)^n}}
 \int_{t_1}^{t_2} d \tau_1 \cdots  \int_{t_1}^{t_2} d \tau_n 
 \ {1 \over \tau_2} \  g_2'\left({\tau_1 \over \tau_2} \right)
\ \prod_{i=2}^n  \ { 1 \over \tau_{i+1}} \  g_0'\left({\tau_i \over \tau_{i+1}} \right)
\ee
\be
=-\sum_{n=1}^{\infty} {{(-2\mu)^n}}
 \int_{\ln t_1}^{\ln t_2} d u_1 \cdots  \int_{\ln t_1}^{\ln t_2} d u_n 
  \  g_2'\left( e^{\displaystyle u_1-u_2} \right)
\ \prod_{i=2}^n  \  g_0'\left( e^{\displaystyle u_i-u_{i+1}} \right)
\ee
The leading behavior of order $ \left ( \ln {t_2 \over t_1}\right)$
may thus be obtained as in \cite{DHP2}
\be
 {\cal T}_{2}(t_1,t_2,q) \simeq -\ln \left ({t_2 \over t_1}\right)
\sum_{n=1}^{\infty} {{(-2\mu)^n}}
\int_{-\infty}^{+\infty} d v_1 \cdots \int_{-\infty}^{+\infty} d v_n
 \ g_2'\left( e^{\displaystyle v_1} \right)
\prod_{i=2}^n   g_0'\left( e^{\displaystyle v_i} \right)
\delta \big(v_1+v_2+ \cdots v_n \big)
\ee
Setting
\be
\delta \big(v_1+v_2+ \cdots v_n \big) =  e^{\displaystyle v_1+v_2+ \cdots v_n}
\int_{-\infty}^{+\infty} {dk \over {2\pi}} 
 e^{\displaystyle ik\big(v_1+v_2+ \cdots v_n \big)} 
\ee
we obtain
\be
 {\cal T}_{2}(t_1,t_2,q) \simeq -\ln \left ({t_2 \over t_1}\right)
\int_{-\infty}^{+\infty} {dk \over {2\pi}} 
\bigg[\int_{-\infty}^{+\infty} d v \  e^{\displaystyle ikv} \ e^v \ g_2'(e^v) \bigg]
\sum_{n=1}^{\infty} {{(-2\mu)^n}}
\bigg[\int_{-\infty}^{+\infty} d w \  e^{\displaystyle ikw} \ e^w \ g_0'(e^w) \bigg]^{n-1}
\ee
\be
 \simeq 2 \mu \ln \left ({t_2 \over t_1}\right)
\int_{-\infty}^{+\infty} {dk \over {2\pi}} 
{I_2(k) \over {1+2\mu I_0(k)}}
\ee
where
\be
I_0(k)=\int_{-\infty}^{+\infty} d w \  e^{\displaystyle ikw} \ e^w \ g_0'(e^w) =
-{2 \over \pi} \int_{-\infty}^{+\infty} d w \  e^{\displaystyle ikw} \ 
{ { e^{\displaystyle w \over 2}} \over {1+e^w}} = -{2 \over {\cosh (\pi k)}}
\ee
and
\be
I_2(k)=\int_{-\infty}^{+\infty} d v \  e^{\displaystyle ikv} \ e^v \ g_2'(e^v)
= -ik \int_{-\infty}^{+\infty} d v \  e^{\displaystyle ikv} \ g_2(e^v)
\ee
Using (\ref{g2z}), we get
\be
I_2(k)=-i k {2 \over {\pi^2}}
 \int_0^1 dx {{e^{-ik \ln x}} \over {\sqrt{x(1-x)}}} 
\int_0^{1} dy  {{e^{ik \ln y}} \over {\sqrt{y(1-y)}}} 
\int_{-\infty}^{+\infty} d w \  e^{\displaystyle ikw} 
 {\min (e^w,1)}  \ {{1-e^w} \over {(1+e^w)^2}}
\ee
The integrals over the variables $x$ and $y$ give
\be
 \int_0^1 dx {{e^{-ik \ln x}} \over {\sqrt{x(1-x)}}} 
\int_0^{1} dy  {{e^{ik \ln y}} \over {\sqrt{y(1-y)}}} 
=\pi {{\Gamma\left({1 \over 2}-ik\right)} \over {\Gamma(1-ik)}}
{{\Gamma\left({1 \over 2}+ik\right)} \over {\Gamma(1+ik)}}
={{\pi \sinh(\pi k)} \over {k \cosh(\pi k)}}
\ee
and the integral over the variable $w$ may be rewritten as
\be
\int_{-\infty}^{+\infty} d w \  e^{\displaystyle ikw} 
 {\min (e^w,1)}  \ {{1-e^w} \over {(1+e^w)^2}}
=-2i \int_0^{\infty} du \sin(ku) {{e^u-1} \over {(e^u+1)^2}}
\ee
to obtain
\be
I_2(k)=-{4 \over {\pi}} \tanh(\pi k) 
\int_0^{\infty}
 du \sin(ku) {{e^u-1} \over {(e^u+1)^2}}
\ee
We therefore get
\be
 {\cal T}_{2}(t_1,t_2,q) \simeq  {{\mu } \over {\pi}}
\ln \left ({t_2 \over t_1}\right)
\int_{-\infty}^{+\infty} {dk } 
{{\cosh(\pi k)} \over {\cosh(\pi k)-4\mu}} \ I_2(k)
\ee
\be
\simeq - {{8 \mu } \over {\pi^2}} \ln \left ({t_2 \over t_1}\right)
\int_0^{\infty}
 du  {{e^u-1} \over {(e^u+1)^2}}
\int_{0}^{+\infty} {dk } \sin(ku)
{{\sinh(\pi k)} \over {\cosh(\pi k)-4\mu}}
\ee
The integral over $k$ gives (in the sense of distributions theory)
\be
\int_{0}^{+\infty} {dk } \sin(ku)
{{\sinh(\pi k)} \over {\cosh(\pi k)-4\mu}}
= {{\cosh(u \delta)} \over {\sinh u}}  \qquad \hbox{where} 
\qquad \delta = {1 \over \pi} \arccos(-4\mu) \in (0.5,1)
\label{defdelta}
\ee
so that
\be
 {\cal T}_{2}(t_1,t_2,q) 
\simeq - {{2 \mu } \over {\pi^2}} \ln \left ({t_2 \over t_1}\right)
\int_0^{\infty}  du  \ e^{-{u \over 2}}
{{\cosh(u \delta)} \over {\left(\cosh {u \over 2}\right)^3}}  
\ee

\vskip 2 true cm

\section{ Asymptotic behavior of ${\cal T}_{1}\lowercase{(t_1,t_2,q)}$}

To compute ${\cal T}_{1,1}(t_1,t_2,q)$, we introduce the notation (\ref{c1})
\be
f\left({\tau_1 \over \tau_2},{\tau_a \over \tau_b},
{\tau_2 \over \tau_b} \right)
\equiv { 1 \over \lambda} \ < c_1^{\{ \cal S\}}(\tau_1,\tau_2) \ c_1^{\{ \cal S\}}(\tau_a,\tau_b)> 
\ee
\be \ba
= \displaystyle {8 \over {\pi^3}} \int_0^1 {dx \over {\sqrt{x(1-x)}}} \ 
\int_0^1 {dy \over {\sqrt{y(1-y)}}} \ 
\bigg[ { {\sqrt{\tau_2 \tau_b} \min(\tau_1 x, \tau_a y)} 
\over {(\tau_2+\tau_1 x)(\tau_b+\tau_a x) }}
+ { {\sqrt{\tau_1 \tau_a} \min(\tau_2 x, \tau_b y)} 
\over {(\tau_1+\tau_2 x)(\tau_a+\tau_b x) } }
\\  \\
- \displaystyle { {\sqrt{\tau_1 \tau_b} \min(\tau_2 x, \tau_a y)} 
\over {(\tau_1+\tau_2 x)(\tau_b+\tau_a x) }}
- { {\sqrt{\tau_2 \tau_a} \min(\tau_1 x, \tau_b y)} 
\over {(\tau_2+\tau_1 x)(\tau_a+\tau_b x) }} \bigg]
\ea \ee
and rewrite
\be
 {\cal T}_{1,1}(t_1,t_2,q)=-\sum_{n=2}^{\infty} {{(-2\mu)^n}\over 2} \sum_{l=2}^n
 \int_{t_1}^{t_2} d \tau_1 \cdots  \int_{t_1}^{t_2} d \tau_n 
 {1 \over {\tau_2 \tau_{l+1}}}  \partial_1 \partial_2 
f\left({\tau_1 \over \tau_2},{\tau_l \over \tau_{l+1}},
{\tau_2 \over \tau_{l+1}} \right)
\prod_{i \neq 1,l}  {1 \over \tau_{i+1}}  
g_0'\left({\tau_i \over \tau_{i+1}} \right)
\ee
\be
=-\sum_{n=2}^{\infty} {{(-2\mu)^n}\over 2} \sum_{l=2}^n
 \int_{\ln t_1}^{\ln t_2} d u_1 \cdots  \int_{\ln t_1}^{\ln t_2} d u_n 
   \partial_1 \partial_2 
f\left( e^{ u_1-u_2},  e^{ u_l-u_{l+1}},
 e^{u_2-u_{l+1}} \right)
\prod_{i \neq 1,l}   g_0'\left( e^{ u_i-u_{i+1}} \right)
\ee
The leading behavior of order $ \left ( \ln {t_2 \over t_1}\right)$
 therefore reads
\be \ba
 {\cal T}_{1,1}(t_1,t_2,q) \simeq -\ln \left ({t_2 \over t_1}\right)
\displaystyle \sum_{n=2}^{\infty} {{(-2\mu)^n}\over 2} \sum_{l=2}^n
\int_{-\infty}^{+\infty} d v_1 \cdots \int_{-\infty}^{+\infty} d v_n
\\   \\
\partial_1 \partial_2 
f\left( e^{v_1},  e^{ v_l},
 e^{ v_2+\cdots+v_l} \right)
\ \displaystyle \prod_{i=2}^n  \  g_0'\left( e^{ v_i} \right)
\ \delta \big(v_1+v_2+ \cdots v_n \big)
\ea \ee
Setting
\be
\delta \big(v_1+v_2+ \cdots v_n \big) =  e^{\displaystyle v_1+v_2+ \cdots v_n}
\int_{-\infty}^{+\infty} {dk \over {2\pi}} 
 e^{\displaystyle ik\big(v_1+v_2+ \cdots v_n \big)} 
\ee
and
\be
1= \int_{-\infty}^{+\infty} dz \ \delta\bigg(z- (v_2+\cdots+v_l) \bigg)
= \int_{-\infty}^{+\infty} dz \ \int_{-\infty}^{+\infty} {dp \over {2\pi}} 
 e^{\displaystyle i(p-k)\big((v_2+ \cdots v_l)-z \big)} 
\ee
we obtain
\be
 {\cal T}_{1,1}(t_1,t_2,q) \simeq - \ln \left ({t_2 \over t_1}\right)
\int_{-\infty}^{+\infty} {dk \over {2\pi}} 
\int_{-\infty}^{+\infty} {dp \over {2\pi}} J(k,p)
\sum_{n=2}^{\infty} {{(-2\mu)^n}\over 2} \sum_{l=2}^n
\bigg[I_0(k) \bigg]^{n-l} \ \bigg[I_0(p) \bigg]^{l-2}
\ee
\be
 \simeq -2 \mu^2 \ln \left ({t_2 \over t_1}\right)
\int_{-\infty}^{+\infty} {dk \over {2\pi}} 
\int_{-\infty}^{+\infty} {dp \over {2\pi}} 
{J(k,p) \over {\bigg[1+2\mu I_0(k)\bigg] \ \bigg[1+2\mu I_0(p)\bigg]}}
\ee
where
\be
J(k,p)=\int_{-\infty}^{+\infty} d u \int_{-\infty}^{+\infty} d v
\int_{-\infty}^{+\infty} d z \  e^{\displaystyle iku+ipv+i(k-p)z} \  e^{\displaystyle u+v} \  
\partial_1 \partial_2 
f\left( e^{\displaystyle u},  e^{\displaystyle v},
 e^{\displaystyle z} \right)
\ee
Since
\be \ba
f\left(e^u,e^v, e^z \right)
= \displaystyle {8 \over {\pi^3}} \int_0^1 {dx \over {\sqrt{x(1-x)}}} \ 
\int_0^1 {dy \over {\sqrt{y(1-y)}}} \ 
\bigg[ { {e^{z \over 2} \min(e^u x, e^{v-z} y)} 
\over {(1+e^u x)(1+e^v x) }}
\\ \\
+ \displaystyle { { e^{z \over 2} \min( x, e^{-z} y)} 
\over {(e^{u \over 2}+e^{-{u \over 2}} x)  
(e^{v \over 2}+e^{-{v \over 2}} y)} }
-  { {e^{z \over 2} \min( x, e^{v-z} y)} 
\over {(e^{u \over 2}+e^{-{u \over 2}} x)(1+e^v x) }}
- { {e^{z \over 2} \min(e^u x, e^{-z} y)} 
\over {(1+e^u x)(e^{v \over 2}+e^{-{v \over 2}} y) }} \bigg]
\ea \ee
it is convenient to begin with the integration over the variable $z$ 
\be
F\left(e^u,e^v, k-p \right)
\equiv \int_{-\infty}^{+\infty} d z \  e^{\displaystyle i(k-p)z} \
f\left( e^{\displaystyle u},  e^{\displaystyle v},
 e^{\displaystyle z} \right)
\ee
and to compute $J(k,p)$ through
\be
J(k,p)= - k p \int_{-\infty}^{+\infty} d u \int_{-\infty}^{+\infty} d v
\  e^{\displaystyle iku+ipv} \  \  
F\left( e^{\displaystyle u},  e^{\displaystyle v}, k-p \right)
\ee
Using successively
\be
\int_{-\infty}^{+\infty} d z \  e^{\displaystyle i(k-p)z}  \ 
e^{z \over 2} \ \min(a,b e^{-z}) = \sqrt{a b } \ {{e^{i (k-p) \ln ({b \over a})}} \over {{1 \over 4} +(k-p)^2}}
\ee
\be
\int_{-\infty}^{\infty} du {{e^{i \omega u} } \over { a e^{u \over 2}+b e^{-{u \over 2} }}} = {{\pi e^{i \omega \ln ({b \over a}) } \over {{\sqrt{ab}} \cosh(\pi \omega)}}  \qquad (\hbox{for} \ \ a>0 \ \ \hbox{and} \ \ b>0} )
\ee
\be
\int_0^1 dx {{e^{i k \ln x}} \over {\sqrt{x(1-x)}}}
=\sqrt{\pi} {{\Gamma\left({1 \over 2}+ik\right)} \over {\Gamma(1+ik)}}
\ee
and well known properties of the $\Gamma$ function, we finally get
\be \ba
J(k,p)= \displaystyle { 8 \over {{1 \over 4} +(k-p)^2}} 
\ {1 \over {\cosh(\pi k) \cosh(\pi p)}}
\\ \\
\displaystyle \bigg[ {{k } \over {\cosh(\pi k)}} \ \sinh(\pi p)+ 
\sinh(\pi k) \ {{p } \over {\cosh(\pi p)}} 
 - k \ p \ 
 \bigg( {{\Gamma \left({1 \over 2}-ik\right)} \over  {  \Gamma(1-ik)}} 
\ {{\Gamma\left({1 \over 2}-ip\right)} \over  {\Gamma(1-ip)}} + c.c.\bigg) \bigg]
\label{Jexpl}
\ea \ee
so that
\be
 {\cal T}_{1,1}(t_1,t_2,q)
 \simeq - { \mu^2 \over {2 \pi^2}} \ln \left ({t_2 \over t_1}\right)
\int_{-\infty}^{+\infty} {dk } 
\int_{-\infty}^{+\infty} {dp} 
{{\cosh(\pi k)} \over {\cosh(\pi k)-4\mu}} 
\ {{\cosh(\pi p)} \over {\cosh(\pi p)-4\mu}} 
\ J(k,p)
\ee
It is convenient to decouple the integrations over the variable 
$k$ and $p$ by setting 
\be
 { 1 \over {{1 \over 4} +(k-p)^2}} = \int_{-\infty}^{+\infty} dx 
\ e^{- {{\vert x \vert} \over 2}} \ e^{i (k-p) x} 
\ee
in (\ref{Jexpl}) to obtain 
\be \ba
\displaystyle {\cal T}_{1,1}(t_1,t_2,q)
\simeq  - 4 {\mu^2 \over {\pi^2}} \ln \left ({t_2 \over t_1}\right)
 \int_{-\infty}^{+\infty} dx \ e^{-{{\vert x \vert} \over 2}} \ 
\int_{-\infty}^{+\infty} {dk } 
\int_{-\infty}^{+\infty} {dp } 
\ \ {e^{ikx} \over {\cosh(\pi k)-4\mu}} 
\ \ {e^{-ipx} \over {\cosh(\pi p)-4\mu}}   
\\ \\ 
\displaystyle 
\bigg[ {{k } \over {\cosh(\pi k)}} \ \sinh(\pi p)+ 
\sinh(\pi k) \ {{p } \over {\cosh(\pi p)}} 
- k p
 \left( {{\Gamma \left({1 \over 2}-ik\right)} \over  {  \Gamma(1-ik)}} 
\ {{\Gamma\left({1 \over 2}-ip\right)} \over  {\Gamma(1-ip)}} + c.c.\right) \bigg]
\ea \ee
Using again (\ref{defdelta}) and
\be \ba
\displaystyle\int_{0}^{+\infty} {dk } \sin(kx)
{{k} \over {\cosh(\pi k) (\cosh(\pi k)-4\mu)}}
= {1 \over {4 \mu}} {{\partial} \over {\partial x}}
\int_{0}^{+\infty} dk  \cos(kx)
\left( {1 \over {\cosh(\pi k)}} - {1 \over { \cosh(\pi k)-4\mu} } \right)
\\ \\
\displaystyle
= {1 \over {4 \mu}} {{\partial} \over {\partial x}}
\left( {1 \over {2 \cosh( {x \over 2})}} - 
{{\sinh(\delta x)} \over { \sin(\pi \delta) \sinh x}} \right)
\ea \ee
leads to
\be \ba
\displaystyle {\cal T}_{1,1}(t_1,t_2,q)
\simeq  8 {\mu \over {\pi^2}} \ln \left ({t_2 \over t_1}\right)
\bigg[{1 \over 4} \int_0^{\infty}  dx  \ e^{-{x \over 2}}
{{\cosh( \delta x)} \over {\cosh^3 {x \over 2}}}  
\\  \\
\displaystyle +{1 \over {\sin (\pi \delta)}}
\int_0^{\infty}  dx  \ e^{-{x \over 2}}
\left( { 2 \delta} {{\cosh^2( \delta x)} \over {\sinh^2 {x}}} 
-{{\sinh( 2 \delta x) \cosh x} \over {\sinh^3 {x }}} \right) \bigg] 
\\  \\
\displaystyle - 16 {\mu^2 \over {\pi^2}} \ln \left ({t_2 \over t_1}\right)
\int_0^{\infty}  dx  \ e^{-{x \over 2}} \  h(x) h(-x) 
\ea \ee
with
\be
h(x) \equiv -i \int_{-\infty}^{+\infty} dk \ 
{{k e^{ikx}} \over {\cosh(\pi k)-4\mu}} \  
{{\Gamma \left({1 \over 2}-ik\right)} \over  {  \Gamma(1-ik)}} 
\ee
\be
= -{{2} \over {\sqrt{\pi} \sin(\pi \delta)}} \ {\partial \over {\partial x}} \ \int_0^{\infty} dz {1 \over {\sqrt{e^z-1}}} \ 
{ {\sinh \delta(x+z)} \over {\sinh(x+z)}}
 \ee
For $x>0$, $h(x)$ may be easily expanded as
\be 
h(x)={{2} \over { \sin(\pi \delta)}} \sum_{n=0}^{\infty}
\left[ e^{-(2n+1-\delta)x} \ {{\Gamma \left(2n+{3 \over 2}-\delta\right)} 
\over {\Gamma \left(2n+1-\delta\right)}}
- e^{-(2n+1+\delta)x} \ {{\Gamma \left(2n+{3 \over 2}+\delta\right)} 
\over {\Gamma \left(2n+1+\delta\right)}} \right]
\ee
and may be rewritten in terms of hypergeometric functions $F(a,b,c,z)$ (\cite{Grad})
\be \ba
h(x)=\displaystyle  {1 \over { \sin(\pi \delta)}} \ \bigg[ e^{-(1-\delta)x}
{{\Gamma \left({3 \over 2}-\delta\right)} \over {\Gamma \left(1-\delta\right)}}
\bigg[ F\left(1,{3 \over 2}-\delta, 1-\delta, e^{-x} \right)
+F \left( 1,{3 \over 2}-\delta, 1-\delta, - e^{-x} \right) \bigg]
\\  \\
\displaystyle - e^{-(1+\delta)x}
{ {\Gamma \left({3 \over 2}+\delta\right)} \over {\Gamma \left(1+\delta\right)}}
\bigg[ F\left(1,{3 \over 2}+\delta, 1+\delta, e^{-x} \right)
+F\left(1,{3 \over 2}+\delta, 1+\delta, - e^{-x} \right) \bigg]\bigg]
\label{hhyper}
\ea \ee 
The analytic continuation of the hypergeometric functions (\cite{Grad}) gives the following expansion for 
$h(-x)$ in the domain $x>0$
\be \ba
h(- x)= \displaystyle  {{2} \over { \cos(\pi \delta)}} 
\bigg( \sum_{n=0}^{\infty}
\left[ e^{-(2n+1-\delta)x} \ {{\Gamma \left(2n+2-\delta\right)} 
\over {\Gamma \left(2n+{3 \over 2}-\delta\right)}}
+ e^{-(2n+1+\delta)x} \ {{\Gamma \left(2n+2+\delta\right)} 
\over {\Gamma \left(2n+{3 \over 2}+\delta\right)}} \right]
\\  \\
\displaystyle - \sum_{l=0}^{\infty} e^{- \left (l+{1 \over 2}\right)x} \  
{{\Gamma \left(l+{3 \over 2} \right)} \over {l!}} \bigg)
\ea \ee

\vskip 2 true cm

\section{ Expression of $\lowercase{\phi(q)}$}

The asymptotic behaviors of ${\cal T}_{2}(t_1,t_2,q)$
and ${\cal T}_{1}(t_1,t_2,q)$ obtained respectively in Appendix B and Appendix C
 imply that $\phi(q)$ defined in (\ref{phiq}-\ref{asphi}) is given for $1<q<2$ in terms of
$\delta= {1 \over \pi} \arccos(-4\mu) ={1 \over \pi} \arccos(-4{{q-1} \over q^2})\in (0.5,1)$ as
\be \ba
\displaystyle \phi(q)=
  {{2 \cos (\pi \delta)} \over {\pi^2 \sin (\pi \delta)}}
\int_0^{\infty}  dx  \ e^{-{x \over 2}}
\left( { 2 \delta} {{\cosh^2( \delta x)} \over {\sinh^2 {x}}} 
-{{\sinh( 2 \delta x) \cosh x} \over {\sinh^3 {x }}} \right) 
\\  \\
\displaystyle +  {{\cos^2(\pi \delta)}\over {\pi^2}}
\int_0^{\infty}  dx  \ e^{-{x \over 2}} \  h(x) h(-x) 
\label{phiint}
\ea \ee
where the function $h(x)$ is given in (\ref{hhyper}) of Appendix C.

The first integral of the formula (\ref{phiint}) can be computed in terms of the logarithmic derivative $\Psi(x) \equiv { {\Gamma'(x)}\over {\Gamma(x)}}$
of the Gamma function
\be \ba
\displaystyle \int_0^{\infty}  dx  \ e^{-{x \over 2}}
\left( { 2 \delta} {{\cosh^2( \delta x)} \over {\sinh^2 {x}}} 
-{{\sinh( 2 \delta x) \cosh x} \over {\sinh^3 {x }}} \right) 
\\ \\
\displaystyle =\int_0^1 dz { {z^{1/4}} \over {(1-z)^3}} \left[ 
\delta (1-z) \left(z^{\delta}+z^{-\delta}+2\right) 
+(1+z) \left(z^{\delta}-z^{-\delta}\right) \right]
\\ \\
\displaystyle ={3 \over 2} \delta+{\delta \over 2} \Psi\left({1\over 4}\right)
+{1 \over 4} \left[ \left({1 \over 4}-\delta\right) \Psi\left({1 \over 4}-\delta\right) -\left({1 \over 4}+\delta\right) \Psi\left({1 \over 4}+\delta\right) \right]
 \ea \ee

To obtain the analytical continuation of $\phi(q)$ to the domain $q\geq 2$,
we need to redefine $\delta$ as
\be
\delta(q) ={1 \over \pi} \arccos\left(-4{{q-1} \over q^2}\right) \qquad 
\longrightarrow \qquad \hat \delta(q)={2 \over \pi} 
\arccos\left({ {2-q} \over {{\sqrt 2 } q}}\right)
\ee
and to rewrite (\ref{phiint}) in a form valid in the whole domain 
$\hat \delta(q=1)= 1/2 <\hat \delta <\hat \delta(q=\infty)=3/2$
\be \ba
\displaystyle \phi(q)=
  {{ \cos (\pi \hat \delta)} \over {\pi^2 \sin (\pi \hat \delta)}}
\left(1+{3 } \hat \delta+{\hat \delta} \Psi\left({1\over 4}\right)
+{1 \over 2} \left[ \left({1 \over 4}-\hat \delta\right) \Psi\left({9 \over 4}-\hat \delta\right) -\left({1 \over 4}+\hat \delta\right) \Psi\left({1 \over 4}+\hat \delta\right) \right]
 \right)
\\  \\
\displaystyle +  {1\over {\pi^2}}
\int_0^1  {dz \over {\sqrt z}} 
\left[ H_+(z,\hat \delta) H_-(z,\hat \delta)
+2 z^{(1-\hat \delta)} 
\left( {{\Gamma(2-\hat \delta)} \over {\Gamma({3 \over 2}-\hat \delta)}} H_+(z, \hat\delta)
-  \left(\hat \delta-{1 \over 2} \right)
 {{\Gamma(\hat \delta)} \over {\Gamma({1 \over 2}+\hat \delta)}} H_-(z, \hat \delta)
\right) \right] 
\ea \ee
where
\be 
H_+(z, \hat \delta)=
{{2\cos(\pi \hat\delta)} \over { \sin(\pi \hat\delta)}} \left[
\sum_{n=1}^{\infty}
 z^{(2n+1-\hat\delta)} \ {{\Gamma \left(2n+{3 \over 2}-\hat\delta\right)} 
\over {\Gamma \left(2n+1-\hat\delta\right)}}
-\sum_{n=0}^{\infty}
 z^{(2n+1+\hat\delta)} \ {{\Gamma \left(2n+{3 \over 2}+\hat\delta\right)} 
\over {\Gamma \left(2n+1+\hat\delta\right)}} \right]
\label{Hplus}
 \ee 
and
\be \ba
\displaystyle H_-(z, \hat \delta)
=2  \bigg[ \sum_{n=1}^{\infty}
 z^{(2n+1-\hat\delta)} \ {{\Gamma \left(2n+2-\hat\delta\right)} 
\over {\Gamma \left(2n+{3 \over 2}-\hat\delta\right)}}
+\sum_{n=0}^{\infty}
 z^{(2n+1+\hat\delta)} \ {{\Gamma \left(2n+2+\hat\delta\right)} 
\over {\Gamma \left(2n+{3 \over 2}+\hat\delta\right)}}
\\  \\
\displaystyle  - \sum_{l=0}^{\infty} z^{ \left (l+{1 \over 2}\right)} \  
{{\Gamma \left(l+{3 \over 2} \right)} \over {l!}} \bigg) \bigg]
\label{Hmoins}
\ea  \ee 

\vskip 2 true cm

\end{document}